\documentclass[
reprint,amsmath,amssymb,prl,superscriptaddress,longbibiliography]{revtex4-2}
\usepackage[version=4]{mhchem}
\usepackage{graphicx,graphics,epsfig}
\usepackage{amsmath, amssymb,amsfonts} 
\usepackage{bm,times,xspace,mhchem}  
\usepackage{mathrsfs}
\usepackage{color}
\usepackage[colorlinks, citecolor=blue, linkcolor=red, citecolor=blue, urlcolor=blue]{hyperref}
\usepackage{dcolumn}
\usepackage{float}
\usepackage{braket}
\usepackage{verbatim}
\usepackage{soul}
\usepackage {lineno}
\begin{document}

\title{Machine Learning of Knot Topology in Non-Hermitian Band Braids}

\author{Jiangzhi Chen}
\altaffiliation{These authors contributed equally to this work}
\affiliation{Center for Phononics and Thermal Energy Science, China-EU Joint Lab on Nanophononics, Shanghai Key Laboratory of Special Artificial Microstructure Materials and Technology, School of Physics Science and Engineering, Tongji University, Shanghai 200092, China}

\author{Zi Wang}
\altaffiliation{These authors contributed equally to this work}
\affiliation{Center for Phononics and Thermal Energy Science, China-EU Joint Lab on Nanophononics, Shanghai Key Laboratory of Special Artificial Microstructure Materials and Technology, School of Physics Science and Engineering, Tongji University, Shanghai 200092, China}

\author{Yu-Tao Tan}
\affiliation{Center for Phononics and Thermal Energy Science, China-EU Joint Lab on Nanophononics, Shanghai Key Laboratory of Special Artificial Microstructure Materials and Technology, School of Physics Science and Engineering, Tongji University, Shanghai 200092, China}

\author{Ce Wang}
\email{
Corresponding address, Ce Wang: phywangce@gmail.com}
\affiliation{Center for Phononics and Thermal Energy Science, China-EU Joint Lab on Nanophononics, Shanghai Key Laboratory of Special Artificial Microstructure Materials and Technology, School of Physics Science and Engineering, Tongji University, Shanghai 200092, China}

\author{Jie Ren}
\email{
Corresponding address, Jie Ren: Xonics@tongji.edu.cn}
\affiliation{Center for Phononics and Thermal Energy Science, China-EU Joint Lab on Nanophononics, Shanghai Key Laboratory of Special Artificial Microstructure Materials and Technology, School of Physics Science and Engineering, Tongji University, Shanghai 200092, China}
\affiliation{Shanghai Research Institute for Intelligent Autonomous Systems, Tongji University, Shanghai 200092, P. R. China}

\begin{abstract}
The deep connection among braids, knots and topological physics has provided valuable insights into studying topological states in various physical systems. However, identifying distinct braid groups and knot topology embedded in non-Hermitian systems is challenging and requires significant efforts. Here, we demonstrate that an unsupervised learning with the representation basis of $su(n)$ Lie algebra on $n$-fold extended non-Hermitian bands can fully classify braid group and knot topology therein, without requiring any prior mathematical knowledge or any pre-defined topological invariants. 
We demonstrate that the approach successfully identifies different topological elements, such as unlink, unknot, Hopf link, Solomon ring, trefoil, and so on, by employing generalized Gell-Mann matrices in non-Hermitian models with $n$=2 and $n$=3 energy bands. 
Moreover, since eigenstate information of non-Hermitian bands is incorporated in addition to eigenvalues, the approach distinguishes the different parity-time symmetry and breaking phases, recognizes the opposite chirality of braids and knots, and identifies out distinct topological phases that were overlooked before. 
Our study shows significant potential of machine learning in classification of knots, braid groups, and non-Hermitian topological phases.


\end{abstract}

\maketitle

\emph{Introduction}-Knot theory holds special significance in science~\cite{atiyah_1990}, which provides a crucial mathematical language to understand the topological properties and interactions of various physical systems~\cite{RN4,RN9,RN14,RN2,RN6,Jones_Polynomial_Yang,wang2017scheme}. 
For example, the knot invariants derived from solutions of the Yang-Baxter equation are used to characterize the entanglement and topological properties of quantum states~\cite{YangCN_yangbaxter,Yang_baxter}. Furthermore, knot theory contributes to the exploration of topological matters, such as topological insulators~\cite{Topological_Chern_vectors} and topological superconductors~\cite{Double_Helix_Superconductor}, which exhibit exotic phases and protected properties. Recently, it has been discovered that the distinct topological phases of the non-Hermitian band structure in reciprocal space can be classified with braid groups, which are closely related to knots~\cite{PhysRevE.87.050101,PhysRevB.96.241404,PhysRevB.101.205417,PhysRevLett.126.010401,RN13,RN5,Patil_nature_2022,DDL_npj_2022,Midya_APL_2023}. Concrete models that can exhibit different braid patterns have been experimentally implemented in platforms such as optical systems~\cite{RN13,RN5,Patil_nature_2022}, quantum circuits~\cite{DDL_npj_2022} and ideal acoustic metamaterials~\cite{RN14}. Band braiding structure is also recently shown to induce topologically protected quantized response~\cite{GongJiangbin_nc_2021}. Closed braids represent knots, which are well-established in mathematics.


 
Yet, there remain great challenges for knot theory, both mathematically and physically in the specific context of non-Hermitian band braids. One of these challenges is the classification and identification of knot types. To overcome the challenge mathematically, the Alexander polynomial was first proposed back in 1923, and subsequently, various other polynomial forms were introduced~\cite{Peter_AM_2004,Jones_polynomial}. However, the algebraic-based methods used to calculate knot invariants are computationally complex. Furthermore, relying solely on one or a few invariants poses challenges in achieving a comprehensive classification of knots. With the rapid development of artificial intelligence, supervised neural networks~\cite{Vandans_pre_2020,DaviesAlex_natrue_2021} are applied to those knot topology. However, this approach relies on tremendous training samples with prior knowledge of known knot types, which necessitates substantial computational resources, and exploring more complex knots remains difficult. 

\begin{figure}[htb!]
  \centering
  \vspace{-2mm}
  \includegraphics[width=\linewidth]{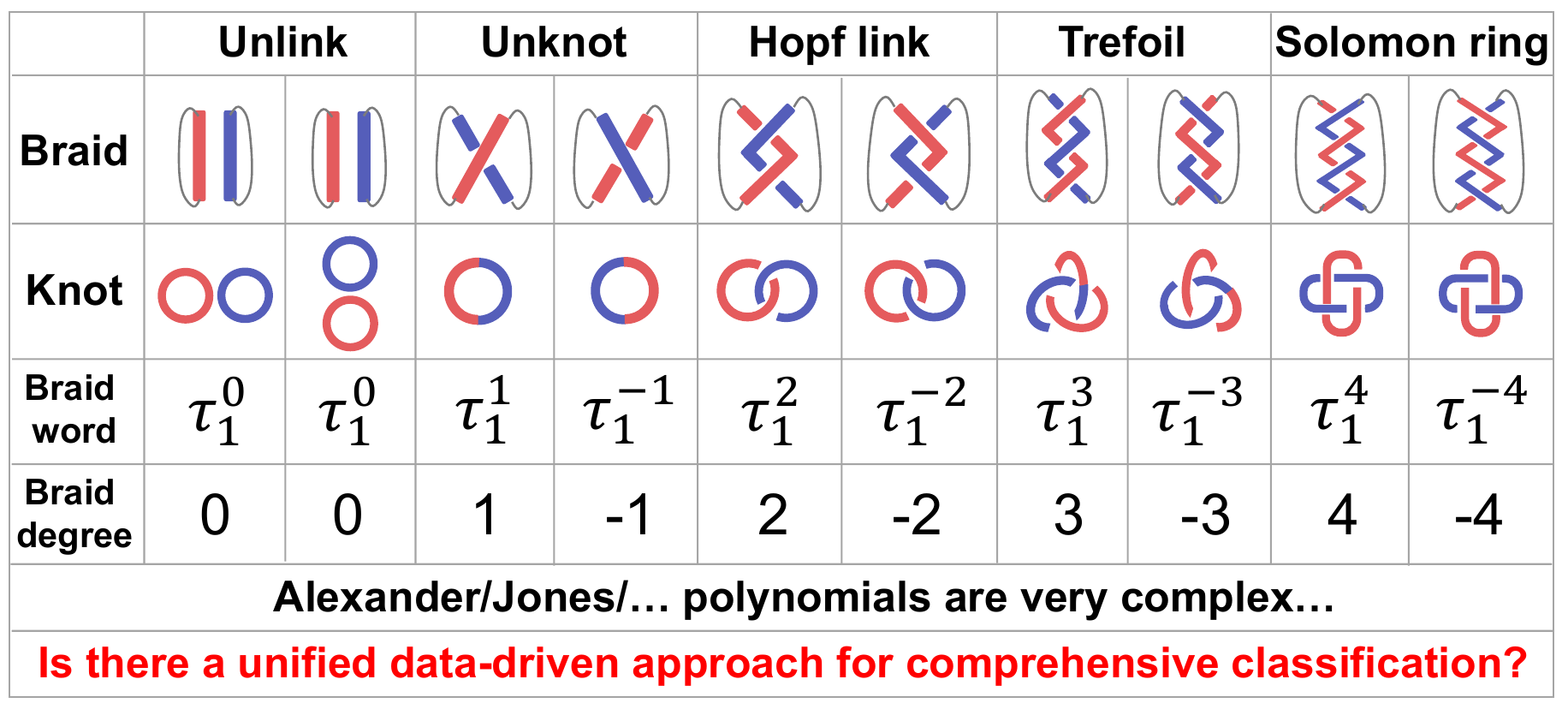}\vspace{-3mm}
  \caption{Examples of braids and knots in $n=2$ non-Hermitian bands with different chiralities and phases. Braid word $\tau_i$($\tau_i^{-1}$) represents the $i$-th string crossing over(under) the ($i+1$)-th string from the left. Braid degree is defined in Eq.~(\ref{invariant}). Complex Alexander and Jones polynomials are not shown. A unified data-driven machine learning classification without any prior mathematical knowledge or any pre-defined topological invariants is highly desired.}
  \label{fig0}
\end{figure}

On the other hand, in the physics context of non-Hermitian band braids where closed braids form rich knot structures, the distinct topological phases are influenced by both complex eigenvalues and eigenvectors~\cite{PhysRevLett.126.010401}. The complex eigenvalues and eigenvectors give rise to intricate phenomena, such as exceptional point (EP), Parity-Time symmetry (PTS) breaking, non-Hermitian skin effect (NHSE). Thus, the topological phase transitions in non-Hermitian systems can be more intricate, manifested not only in the knot topology transitions in band braids, but also in the phase transitions in eigenvectors. As shown in Fig.~\ref{fig0}, the band structure can be restricted by additional symmetries, resulting in different phases with the same knot topology. Furthermore, different chiralities can also result in different band braids. Therefore, a fully classification of non-Hermitian knots and band braids goes beyond the realm of purely mathematical problems, solely based on braid words, braid degrees, and other existing complex methods in knot theory (Fig.~\ref{fig0}). There is an urgent need to develop an efficient and unified approach to identify and classify the rich topological phases of non-Hermitian band systems.

Unsupervised machine learning methods have shown promising capabilities in purely data-driven phase classification~\cite{WangLei_prb_2016,wang2017machine,Wetzel_pre_2017,Rodriguez-Nieva_Natphys_2019,Lidiak_prl_2020,LongYang_prl_2020,Scheurer_prl_2020,CheYanming_prb_2020,YuLiwei_prl_2021,ParkSungjoon_prb_2022,DDL_npj_2022,GongQH_LPR_2023,LongYang_prl_2023}, without requiring prior mathematical knowledge. Specifically, diffusion mapping~\cite{Coifman_pnas_2005,Coifman_acha_2006}, has demonstrated significant advantages in identifying topological phase transitions in various systems, such as the Ising gauge theory~\cite{Rodriguez-Nieva_Natphys_2019,Lidiak_prl_2020}, topological condensed matters~\cite{LongYang_prl_2020,Scheurer_prl_2020,CheYanming_prb_2020,ParkSungjoon_prb_2022}, and non-Hermitian systems~\cite{DDL_npj_2022,YuLiwei_prl_2021,GongQH_LPR_2023}.
The effectiveness of diffusion mapping is due to the natural connection between the diffusion process in sample manifold and the homotopy in topology. However, the performance depends crucially on the definition of diffusion distances between samples.

In this work, we adopt a diffusion distance measure based on the Bloch vector on the basis of $su(n)$ Lie algebra, so called generalized Gell-Mann matrices, to describe non-Hermitian systems with unit cell size $n$. This measure can incorporate information of both complex eigenvalues and eigenstates to fully classify the band braid and knot topology in non-Hermitian systems. Importantly, we note that non-Hermitian band braids in a topologically nontrivial phase may enable the occurrence of spontaneous symmetry breaking, of which each mode does not preserve the $2\pi$ period that the Hamiltonian provides~\cite{PhysRevE.87.050101}. As such, we adopt the momentum space of the descriptors in the $n$-fold extended non-Hermitian Brillouin zone to ensure that each mode remains $2n\pi$ periodicity throughout. 
We take two models as examples, one with $n=2$ and the other with $n=3$. As a result, the diffusion mapping on the measure effectively distinguishes different types and chirality of band braids and knots. It also identifies distinct topological phases with different hidden symmetries even they share the same braid and knot.

\emph{Method}- The traceless part $\hat{\mathbf{H}}^{'}$ of the non-Hermitian $n\times n$ Hamiltonian $\hat{\mathbf{H}}$ can be represented on the basis of $su(n)$ Lie algrebra, so called generalized Gell-Mann matrices~\cite{SM}, which manifests as the Gell-Mann matrices when $n=3$ and reduces to the usual Pauli matrix when $n=2$. For the non-Hermitian system with $n=2$, which is expressed as:
\begin{equation}
\hat{\mathbf{H}}^{'}(k) =\mathbf{h}(k)\cdot \mathbf{\hat{\sigma}}=h_x(k)\hat{\sigma}_x+h_y(k)\hat{\sigma}_y+h_z(k)\hat{\sigma}_z.
\label{trace}
\end{equation}
The descriptor for the $i$-th Hamiltonian sample is identified as the generalized Bloch vector $\mathbf{d}_{\pm}^i(k)\equiv \mathbf{h}^i(k)/E^i_{\pm}(k)$ with $E_{\pm}^2(k)=h_x^2(k)+h_y^2(k)+h_z^2(k)$ the  two continuous complex energy bands. Due to the occurrence of spontaneous symmetry breaking in topologically nontrivial phases, the energy bands no longer exhibit periodicity with a period of 2$\pi$~\cite{PhysRevE.87.050101}. Thus, we expand the momentum space to $k\in(0,2n\pi)$ and ensure $E_{\pm}(k)=E_{\pm}(k+2n\pi)$. 
However, unlike the Hermitian system where the distance between two samples can be defined from the occupied energy band below certain Fermi level, the occupied band for non-Hermitian systems with complex band braiding is not well defined. Thus, we formally introduce the following distance between sample $i$ and $j$ as 
\begin{equation}
  M_{i,j} = \sum_{b\in\{+,-\}}\min_{b'\in\{+,-\}}\left(\left\|\hat{\mathbf{d}}_{b}^i-\hat{\mathbf{d}}_{b'}^j\right\|_{\mathbb{L}_1}^2\right),
\end{equation}
which is the minimum distance between all possible pairs of modes. Discretizing uniformly along the $n$-fold extended non-Hermitian
Brillouin zone $k\in(0,2n\pi)$ with $L$ points, each $\hat{\mathbf{d}}_{b}^i$ is a  $L\times3$ matrix and $\left\|\cdot\right\|$ represents the taxicab $\mathbb{L}_1$-norm distance.
This matrix-based descriptor also finds effective applications in neural network recognition of topological invariants~\cite{ZhaiHui_prl_2018}.

For the higher dimensions systems of $n=3$, the traceless part $\hat{\mathbf{H}}^{'}$ is expressed on the basis of $su(3)$ Lie algebra, as: 
\begin{equation}
\hat{\mathbf{H}}^{'}(k) =\mathbf{h}(k)\cdot \mathbf{\hat{a}}=\sum_{m=1}^{8}h_m(k)\hat{a}_m,
\label{trace3}
\end{equation}
where $\hat{a}_{1,2,..,8}$ are the Gell-Mann matrices. The descriptor for the $i$-th sample is $\mathbf{d}_{1,2,3}^i(k)\equiv \mathbf{h}^i(k)/E^i_{1,2,3}(k)$, where $E^i_{1,2,3}(k)$ are the three continuous complex energy bands with $k\in(0,2n\pi)$. The sample distance is defined as:
\begin{equation}
  M_{i,j} = \sum_{b=1}^{3}\min_{b'\in\{1,2,3\}}\left(\left\|\hat{\mathbf{d}}_{b}^i-\hat{\mathbf{d}}_{b'}^j\right\|_{\mathbb{L}_1}^2\right),
\end{equation}


The similarity matrix is then given by a Gaussian kernel,
\begin{equation}
    K_{i,j}={\rm exp}
    \left(-\frac{M_{i,j}}{2\epsilon L^2}\right),
\label{similarity1}
\end{equation}
where $\epsilon$ is the Gaussian kernel coefficient.
Using Eq.~(\ref{similarity1}), we obtain the similarity matrix of $N$ randomly sampled Hamiltonians in the parameters space, i.e., $K_{i,j}\xrightarrow{}1$ if two samples are similar, otherwise, $K_{i,j}\xrightarrow{}0$. The probability transition matrix is defined as $P_{i,j}=K_{i,j}/\sum_{j=1}^{N}K_{i,j}$ to describe the diffusion progress. After $t$ steps, the diffusion  distance between sample $i$ and $j$ on the manifold is $D^{t}_{i,j}=\sum_{n}[(P_{i,n}^{t}-P_{j,n}^{t})^2/\sum_{n'}K_{n,n'}]=\sum_{n=1}^{N-1}\lambda_{n}^{2t}[(\psi_{n})_i-(\psi_{n})_j]^2$, where $\psi_n$ is the $n$-th right eigenvectors of $\hat{\mathbf{P}}$ and $\lambda_{n}$ is the $n$-th eigenvalue, $n=0,1,...,N-1$. After long time diffusion, the first few components $\psi_n$ that have the largest eigenvalues $\lambda_n\approx1$ will dominate in the manifold diffusion process. Consequently, the prime information of the manifold diffusion distance is encoded within these few components, which reveals the classification information of different phases (see details of the method in ~\cite{SM}).

\begin{figure*}[!htb]
  \centering
  \includegraphics[width=\linewidth]{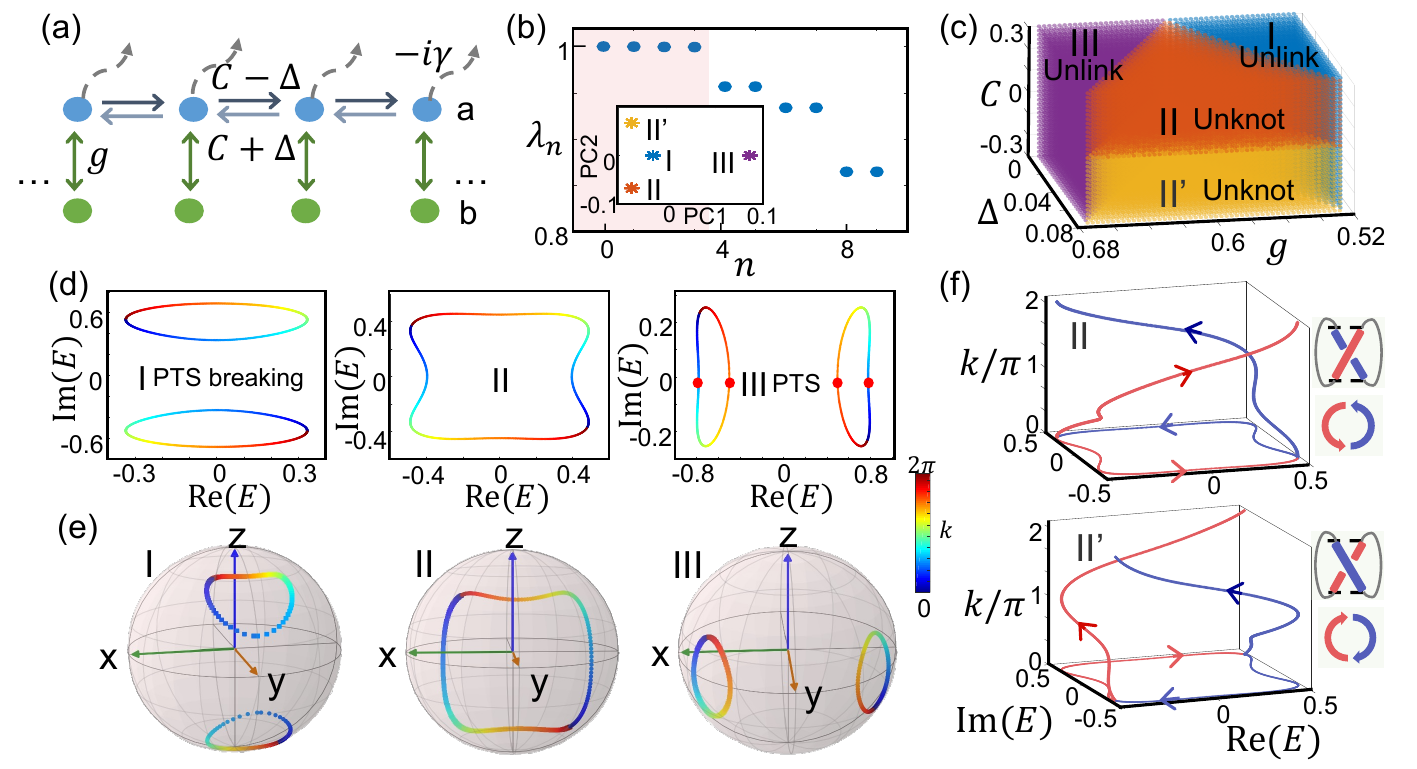}\vspace{-3mm}
  \caption{\textbf{The manifold classification of the $n=2$ non-Hermitian model with asymmetric nearest-neighbor coupling.} (a) Sketch of the two-band lattice model. $C$, $\Delta$, $g$ are the coupling constants. The asymmetric coupling $C\pm\Delta$ is between the nearest-neighbor sites in sublattice $a$. $\gamma>0$ denotes the additional loss rate in sublattice $a$. (b) The first ten largest eigenvalues $\lambda_n$. The  Hamiltonian samples are uniformly generated in the parameter space consisting of variables $\Delta$, $g$ and $C$ with $\gamma=1.2$. The Gaussian kernel coefficient $\epsilon=0.3$. The inset is the first and second principal components perform data dimension reduction on the first four eigenvectors $\psi_n$. The samples are classified into four phases.(c) The phases diagram in the parameter space. (d) The eigenvalues of the traceless Hamiltonian $\hat{\mathbf{H}}'(k)$ and (e) the eigenstates of $\hat{\mathbf{H}}'(k)$ on the Bloch sphere according to phases I (an unlink with PTS breaking), II (an unknot) and III (an unlink with PTS), indicating different topological phases. (f) The interwined complex eigen-energies of the system in phases II and II'.  The scatter plot in the $k=0$ plane represents the projection of the energy bands. The insets illustrate the corresponding braid and knot diagrams with opposite chiralities.
  }
  \label{fig1}
\end{figure*}

\emph{Model and results}- We begin with a concrete $n=2$ model which has been realized experimentally recently by utilizing the frequency modes in two coupled ring resonators~\cite{RN5}. The schematic diagram for this model is shown in Fig.~\ref{fig1}(a), with asymmetric coupling constants $C\pm\Delta$ between the adjacent lattice sites within sublattice $a$. $\gamma>0$ denotes the additional loss rate in sublattice $a$. Each site in sublattice $a$ couples to a site on sublattice $b$ with the strength $g$, and there is no coupling between the sites in sublattice $b$. In momentum-space, the Hamiltonian of this model is written as
\begin{equation}
    \hat{\mathbf{H}}(k)=
    \begin{bmatrix}
    2C{\rm cos} (k)+i 2\Delta {\rm sin} (k)-i\gamma&g\\
    g&0
    \end{bmatrix}.
    \label{hamitonian}
\end{equation}
To explore the versatile phase diagram of this model, we consider a data set with fixed $\gamma=1.2$ and variable $g, \Delta, C$.  
After the unsupervised learning, as illustrated in Fig.~\ref{fig1}(b), we observe the first four eigenvalues  $\lambda_n$ of the probability transition matrix $\hat{\mathbf{P}}$ are much closer to 1 than the rest. Additionally, the visualization of the first four eigenvectors $\psi_{0-3}$~\cite{SM} with the first two principal components is displayed in the inset of Fig.~\ref{fig1}(b). These results indicate the whole data set is classified into four phases autonomously by the algorithm, as shown in Fig.~\ref{fig1}(c). In phases I and III, the non-Hermitian bands form unlinks, whereas in phases II and II', they form unknots.

We compare the topological classification results obtained from unsupervised learning with those obtained using the previously proposed numerical calculation method for braid degree. For a general non-Hermitian Hamiltonian, the braid degree~\cite{GongZongping_prx_2018} $\nu$ can be given by
\begin{equation}
\nu=\int_{0}^{2\pi}\frac{\rm{d}k}{2\pi i}\frac{\rm{d}}{\rm{d}k}\rm{lnDet}(\hat{\mathbf{H}}'(k)),
\label{invariant}
\end{equation}
where $\hat{\mathbf{H}}^{'}(k)$ is the traceless part of the original Hamiltonian. Calculated with Eq.~\ref{invariant}, the braid degrees for phase I and III are both 0, while for phase II and II', they are 1 and -1, respectively. The results obtained from unsupervised learning align with the numerical calculation of the braid degree, but they also provide more comprehensive classification information than the latter.


We analyze the topological properties and knots of different phases by plotting energy bands and eigenvectors of Hamiltonian. Fig.~\ref{fig1}(d) displays the complex eigenvalues $E$ of $\hat{\mathbf{H}}'(k)$ of the phases I, II, III. In phases I and III, the two bands intertwine into two separate rings, forming an unlink, while in phase II, the two bands are connected at their ends, forming a loop, which represents an unknot. Both phases I and III form unlink with the same braid degree $\nu = 0$ according to Eq.~\ref{invariant}, however, they are recognized separately as two distinct topological phases. The qualitative difference between phases I and III is reflected in the energy band structure as demonstrated in Fig.~\ref{fig1}(d). In phase I, we have $E_{\pm}(k) = -E_{\pm}^{*}(\pi-k)$ and the system exhibits PTS breaking. In phase III, however, $E_{\pm}(k) = E_{\pm}^{*}(\pi-k)$. The system exhibits PTS at the high symmetry points $k=\pi/2$ and $k=3\pi/2$. Those relations are protected by the ``hidden'' time-reversal symmetry in this model~\cite{SM,kawabata2019topological}, under which phases I and III cannot deform continuously to each other without band degeneracy. Moreover, the symmetry constraint is similarly reflected in eigenstates for phases I and III. Visualizing the eigenstates as two Bloch vectors $(S_{x}^{\pm},S_{y}^{\pm},S_{z}^{\pm})$  in Fig.~\ref{fig1}(e), we find in phase I, $S_{x}^{\pm}(k)=-S_{x}^{\pm}(\pi-k), S_{y}^{\pm}(k)=S_{y}^{\pm}(\pi-k), S_{z}^{\pm}(k)=S_{z}^{\pm}(\pi-k)$, while different in phase III, $S_{x}^{\pm}(k)=S_{x}^{\pm}(\pi-k), S_{y}^{\pm}(k)=S_{y}^{\pm}(\pi-k), S_{z}^{\pm}(k)=-S_{z}^{\pm}(\pi-k)$. To further demonstrate the differences among phase I, II and III, we plot the Riemann surfaces near EPs in Fig.~\textbf{S2}~\cite{SM}, which are linked to the implementation of the two-band braids.  More specifically, Fig.~\textbf{S3}~\cite{SM} provides a more detailed analysis of the eigenvalues and eigenvectors, revealing distinct behaviors of phases I, II, III before and after the EPs, supporting that they represent three distinct phases. The different topological phases induced by the parameter $g$ exhibit not only distinct eigenvalue behaviors but also different dynamical (eigenstate) behaviors on the sides of EP (Fig.~\textbf{S4}~\cite{SM}).

Fig.~\ref{fig1}(f) displays the complex eigenvalue $E$ of phases II ($\nu=1$) and II' ($\nu=-1$). The scatter plot in the $k=0$ plane represents the knot projection of the energy band braids onto that plane. The insets illustrate the corresponding braid diagrams with braid closures by the grey lines as well as the knot diagrams. Although the band braids of II and II' both form the seemly same unknot, the chirality of the braid is reversed by the sign of $C$,  which is manifested by the direction inversion of the non-Hermitian skin effect (NHSE) under open boundary conditions~\cite{SM}, and is successfully identified as distinct topological phases by the unsupervised machine learning.

\begin{figure}[htb!]
    \centering
    \includegraphics[width=\linewidth]{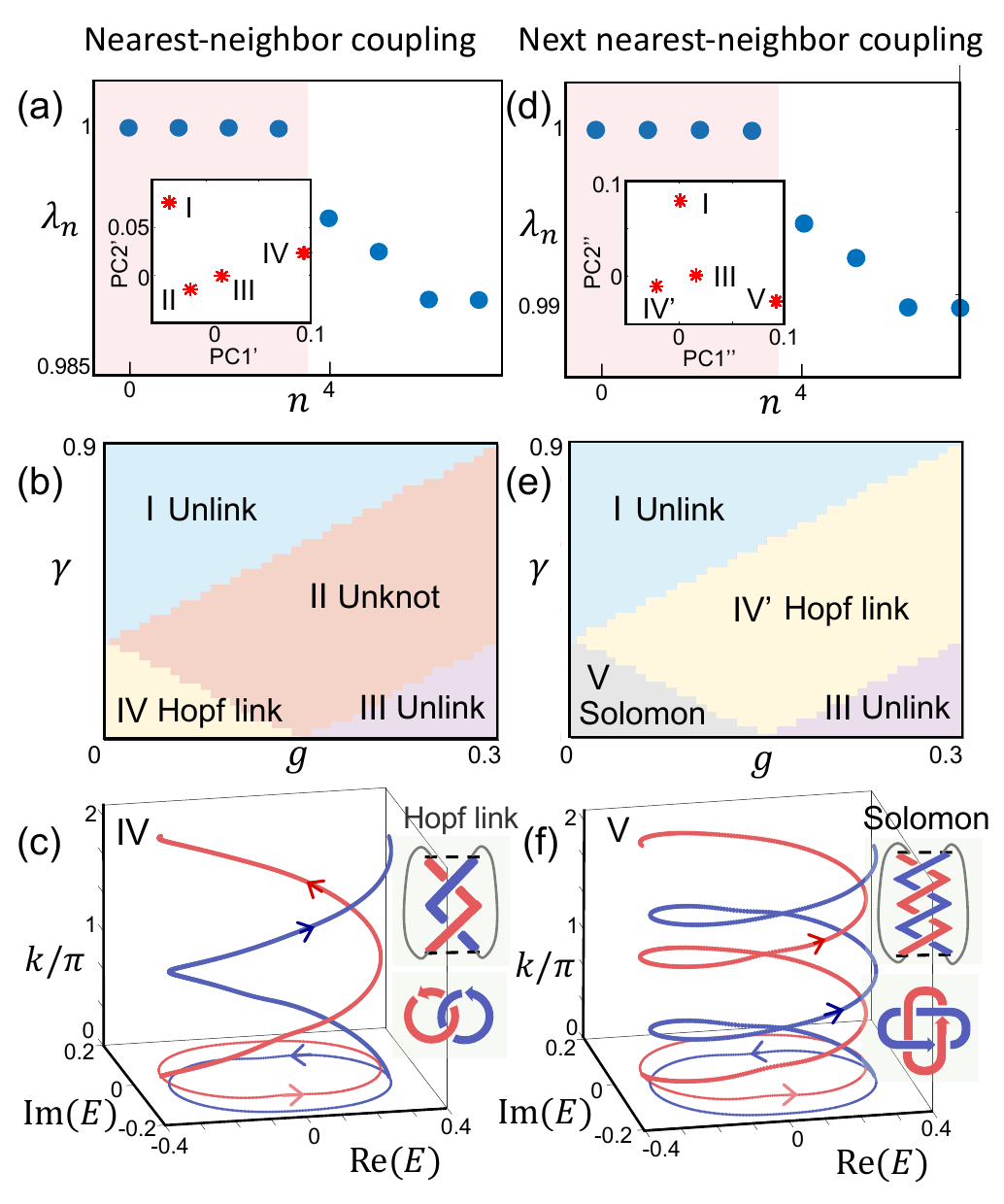}\vspace{-3mm}
    \caption{\textbf{The unsupervised learning classification of $n=2$ non-Hermitian models with (a)-(c) nearest-neighbor and (d)-(f) next nearest-neighbor couplings, varying parameters $g$ and $\gamma$.} (a) The first eight largest eigenvalues $\lambda_n$ and the principal components of the first four eigenvectors $\psi_n$ of the probability transition matrix $\hat{\mathbf{P}}$, indicating four distinct phases. The parameters are $\Delta$=0.15, $C=0.3$. The Gaussian kernel coefficient $\epsilon$=0.001. (b) The phases diagram coincides with the unsupervised learning classification. Apart from the phases \uppercase\expandafter{\romannumeral1}, \uppercase\expandafter{\romannumeral2}, \uppercase\expandafter{\romannumeral3} shown above, a phase \uppercase\expandafter{\romannumeral4} appears when $\gamma<2\Delta-2g$. (c) The interwined complex eigenenergies of the system in phase \uppercase\expandafter{\romannumeral4} forms a Hopf link. (d) The eigenvalues and the corresponding principal components of eigenvectors of the probability transition matrix $\hat{\mathbf{P}}$ for the model with next nearest-neighbor coupling, using the same parameters as in (a). (e) When $\gamma<2\Delta-2g$, the interwined complex eigenenergies of the system braid as a Solomon ring, as shown in (f).}
    \label{fig2}
\end{figure}

Building upon the model illustrated in Fig.~\ref{fig1}(a), we then consider a data set with $\Delta$=0.15, $C$=0.3 and variable $g$, $\gamma$. The classification results of unsupervised learning are shown in Fig.~\ref{fig2}(a), indicating there are four distinct phases. Interestingly, apart from the phases I, II, III mentioned above, a phase IV with $\nu$=2 (calculated from Eq.~\ref{invariant}) appears when $\gamma<2\Delta-2g$, as shown in Fig.~\ref{fig2}(b). Fig.~\ref{fig2}(c) displays the complex eigenenergies of the system in phase IV, which exhibit a Hopf link topology. Similarly, when the asymmetric couplings $C\pm\Delta$ appear in the next-nearest neighboring sublattice $a$ instead of the nearest neighboring sublattice, there are also four distinct phases with varying $g$ and $\gamma$ (Fig.~\ref{fig2}(d)). As depicted in the phase diagram presented in Fig.~\ref{fig2}(e), when $\gamma<2\Delta-2g$, there will be a case of $\nu=4$ (calculated from Eq.~\ref{invariant}) with the interwined band knot corresponding to a Solomon ring. Fig.~\ref{fig2}(f) shows the complex eigenenergies of the system forming Solomon ring.
A similar knot has also been discovered in ferroelectric polarization recently~\cite{WangJing_NC_2023}.  Additionally, when there are lattice sites with further asymmetric couplings, a richer variety of knots can emerge. However, due to limitations in the parameter space, we only display a partial representation of these knots in this study.


\begin{figure}[!htb]
  \centering
  \includegraphics[width=\linewidth]{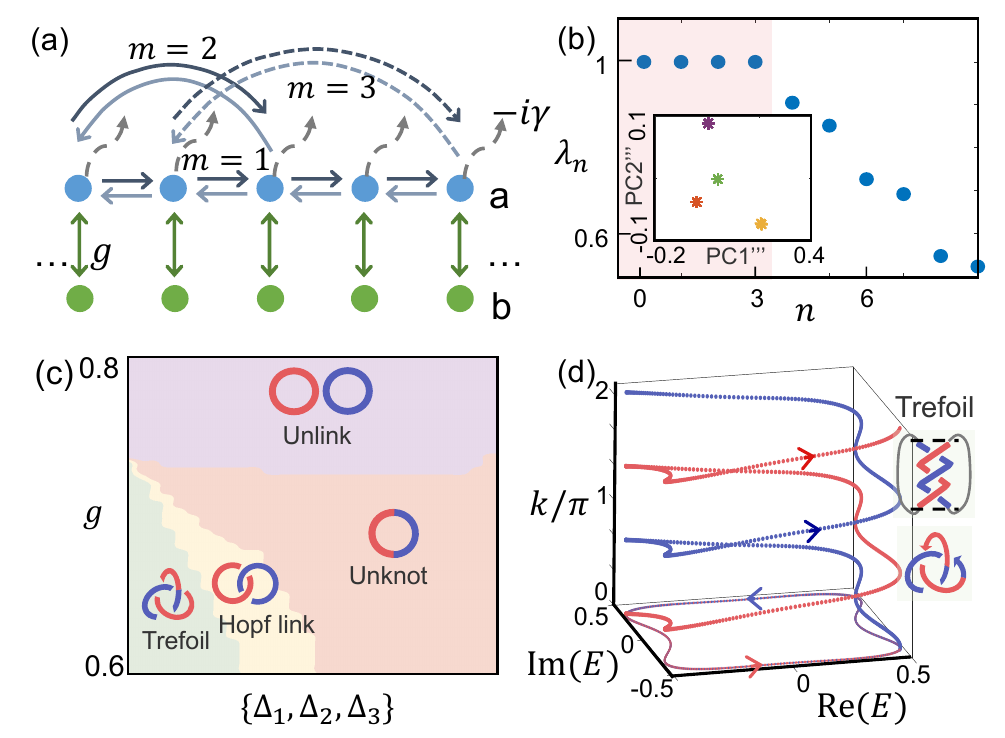}\vspace{-3mm}
  \caption{\textbf{The unsupervised learning classification of a $n=2$ non-Hermitian model with multiple couplings.} (a) Sketch of the two-band lattice model. The asymmetric coupling $C_m\pm\Delta_m$ occurs at the $m$-th ($m$=1,2,3) nearest neighbors. (b) The first ten largest eigenvalues $\lambda_n$. The inset is the principal components of the first four eigenvectors $\psi_n$. The Hamiltonian samples are randomly generated in the parameter space consisting of variable $g$ and $\Delta_{m}$. All the samples are classified into four phases, corresponding to different elements in the braid group. The Gaussian kernel coefficient $\epsilon$=0.05. (c)The phases diagram obtained by the unsupervised learning, with the corresponding knot diagrams and braid degrees. (d)The complex eigen-energies of the system and the corresponding braid and knot diagrams in Trefoil phase. The scatter plot in the $k=0$ plane represents the projection of the energy bands. }
  \label{fig3}
\end{figure}

To evaluate the effectiveness of unsupervised learning in classifying complex braids associated with intricate knots and links, we employ adjustable couplings at the $m$-th ($m$=1,2,3) nearest neighbors, as depicted in Figure~\ref{fig3}(a). The Hamiltonian of this multiple model ($m$=1,2,3) in momentum-space is expressed as:
\begin{equation}
    \hat{\mathbf{H}}(k)=
    \begin{bmatrix}
    \sum_{m}(2C_m{\rm cos} (mk)+i 2\Delta_m {\rm sin} (mk))-i\gamma&g\\
    g&0
    \end{bmatrix}.
    \label{hamitonian}
\end{equation}  
The parameters $\left\{\Delta_m\right\}$ with ($m$=1,2,3) vary simultaneously and continuously while ensuring $C_m=2\Delta_{m}$ (see details on parameters setting in ~\cite{SM}).

Following the unsupervised learning classification, we obtain the eigenvalues $\lambda_n$ in Fig.~\ref{fig3}(b). The first and second principal components of the eigenvectors $\psi_{0-3}$ are shown in the inset, indicating the four distinct phases~\cite{SM}.  Fig.~\ref{fig3}(c) exhibits the well classified phases diagram with the corresponding braid degrees (calculated from Eq.~\ref{invariant}). When $g$ is small, three non-trivial cases of the system's complex energy band emerge as a result of changes in $\left\{\Delta_m\right\}$, corresponding to unknot, Hopf link and Trefoil. When $g$ is large, two bands always form two separate loops, resulting in trivial unlinks. The classification results obtained from unsupervised manifold learning are justified by the distinct braids and knot topology. The corresponding complex-energy bands in phase Trefoil in momentum space are shown in Fig.~\ref{fig3}(d). The complex energy bands of the remaining three phases (unlink, unknot and Hopf link) have been presented in the previous text. Without loss of generality, we only discuss the case $\left\{C_m\right\}>0$ here. As previously mentioned, the chirality of nontrivial phases - unknot, Hopf link, Solomon ring and Trefoil can be inversed by the sign of $\left\{C_m\right\}$, which can also be successfully identified as distinct topological phases by the unsupervised learning~\cite{SM}.

\begin{figure}[!htb]
  \centering
  \includegraphics[width=\linewidth]{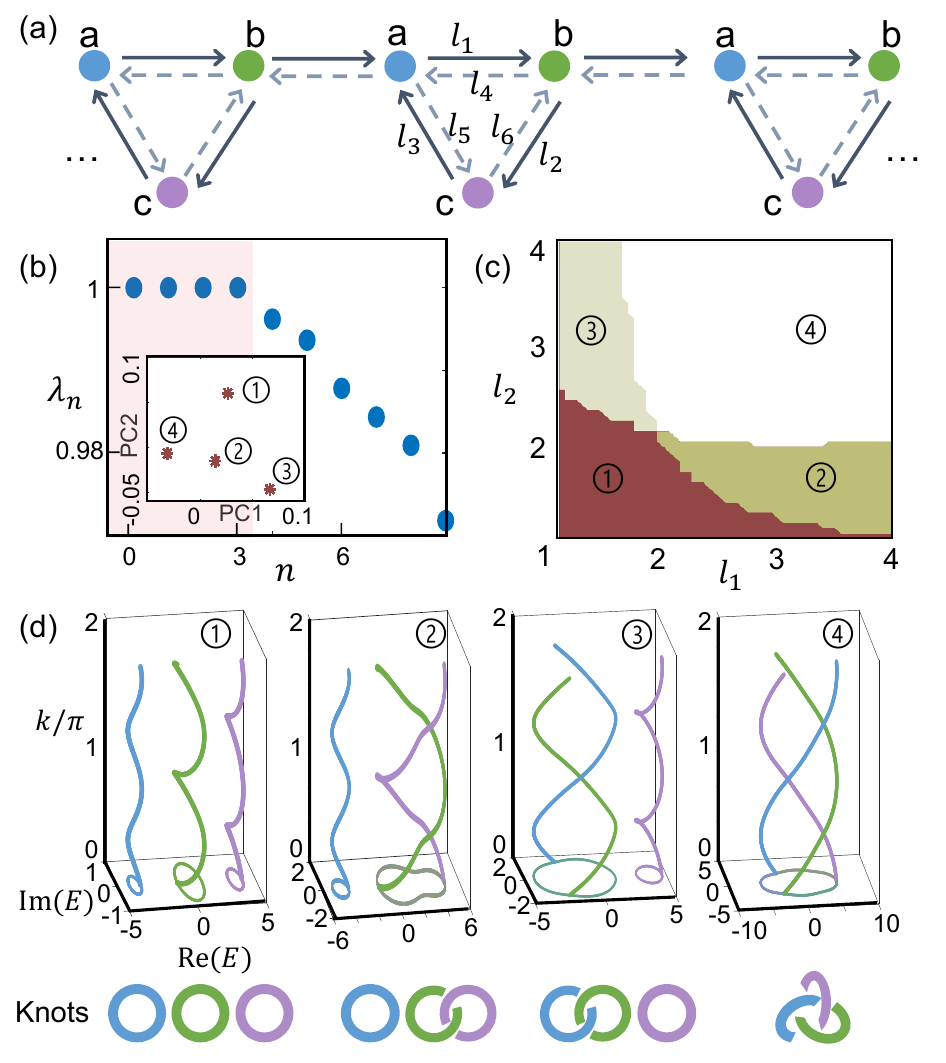}\vspace{-3mm}
  \caption{\textbf{The unsupervised learning classification of a $n=3$ non-Hermitian model.} (a) Sketch of the Markovian three-band kinetic model. $l_1,l_2,l_3,l_4,l_5,l_6$ are the coupling constants. (b) The first ten largest eigenvalues $\lambda_n$ and the principal components of the first four eigenvectors $\psi_n$. The Hamiltonian samples are generated by varying the values of $l_1$ and $l_2$, while keeping $l_3$=5, $l_4$=2, $l_5$=5, and $l_6$=0.1 fixed. These samples are classified into four distinct phases. The Gaussian kernel coefficient $\epsilon$=0.5. (c) The phase diagram in the parameter space obtained through unsupervised learning. (d) The corresponding complex energy bands.}
  \label{fig4}
\end{figure}

We further demonstrate the machine learning of the braid group and knot topology in a $n=3$ non-Hermitian bands. As shown in Fig.~\ref{fig4}(a), we consider a Markovian three-state model which is widely used for counting statistics~\cite{PhysRevE.87.050101}, cyclic enzyme reactions~\cite{Qian_Hong_ARPC_2007}, molecular motors~\cite{Astumian_ARB_2011}, and charge currents through quantum dots~\cite{PhysRevB.81.125331}. The non-Hermitian Hamiltonian is:
\begin{equation}
    \hat{\mathbf{H}}(k)=
    \begin{bmatrix}
    -l_1-l_5&l_{4}e^{-ik}&l_3\\
    l_{1}e^{ik}&-l_2-l_4&l_6\\
    l_5&l_2&-l_3-l_6
    \end{bmatrix}.
    \label{hamitonian2}
\end{equation}  
For the sake of simplicity, we set $l_3$=5, $l_4$=2, $l_5$=5, and $l_6$=0.1, focusing solely on detecting the phase transition in knot topology within the parameter space of ($l_1$, $l_2$). The unsupervised learning results using the proposed method are displayed in Fig.~\ref{fig4}(b). The eigenvalues and eigenvectors of the probability transition matrix indicate the presence of four topological phases. The phases diagram according to the unsupervised learning outcomes is depicted in Fig.~\ref{fig4}(c). Fig.~\ref{fig4}(d) shows the complex energy bands in each phases. In phase \textcircled{1}, the three bands intertwine into three separate rings, forming a trivial unlink with a consistent period of 2$\pi$. As we delve into phase \textcircled{2}, the green and purple bands intertwine through a continuous change of $k$, resulting in a Hopf link, while the blue band forms a separate loop. In phase \textcircled{3}, however, the blue and green bands twist as a Hopf link. Phases \textcircled{2} and \textcircled{3} are evidently distinct, as it is not possible to continuously transform from one phase to the other. Advancing further to phase \textcircled{4}, the three bands twist around each other, forming a Trefoil. Due to the spontaneous symmetry breaking~\cite{PhysRevE.87.050101}, the period of the bands become $6\pi$ rather than $2\pi$. This expansion in period reaffirms the significance of extending the momentum space of the descriptor to $k\in(0,2n\pi)$ within our method, allowing us to capture the properties embedded in the Hamiltonian more comprehensively.

To summarize, we introduced a diffusion distance measure based on the Bloch vector, utilizing the $su(n)$ Lie algebra framework, to characterize one-dimensional non-Hermitian systems with a unit cell size of $n$. Significantly, we extended the momentum space of the descriptors to $k\in(0,2n\pi)$, effectively accommodating the possibility of spontaneous symmetry breaking. Combining the introduced diffusion distance measure with unsupervised learning, we successfully classified the the braid group and knot topology in both $n$=2 and $n$=3 non-Hermitian systems. Furthermore, the method enabled the distinction of the chirality of band braids. It also identified distinct topological phases embedded in eigenstates and protected by hidden symmetry, which were previously overlooked. The proposed method can be extended to other non-Hermitian models, such as non-Hermitian quantum models~\cite{PhysRevLett.130.163001} to detect phases which are currently unknown. Our results prove significant potential of purely data-driven machine learning in uncovering novel insights in the knot topology, braid groups, and non-Hermitian systems, without human knowledge.

\begin{acknowledgments}
This work is supported by National Natural Science Foundation of China (No. 11935010, No. 12204352), the Natural Science Foundation of Shanghai (Nos. 23ZR1481200, 23XD1423800), the National Key R\&D Program of China (Grant No. 2022YFA1404400), and the Opening Project of Shanghai Key Laboratory of Special Artificial Microstructure Materials and Technology.
\end{acknowledgments}

\bibliography{ref}

\begin{thebibliography}{49}%
\makeatletter
\providecommand \@ifxundefined [1]{%
 \@ifx{#1\undefined}
}%
\providecommand \@ifnum [1]{%
 \ifnum #1\expandafter \@firstoftwo
 \else \expandafter \@secondoftwo
 \fi
}%
\providecommand \@ifx [1]{%
 \ifx #1\expandafter \@firstoftwo
 \else \expandafter \@secondoftwo
 \fi
}%
\providecommand \natexlab [1]{#1}%
\providecommand \enquote  [1]{``#1''}%
\providecommand \bibnamefont  [1]{#1}%
\providecommand \bibfnamefont [1]{#1}%
\providecommand \citenamefont [1]{#1}%
\providecommand \href@noop [0]{\@secondoftwo}%
\providecommand \href [0]{\begingroup \@sanitize@url \@href}%
\providecommand \@href[1]{\@@startlink{#1}\@@href}%
\providecommand \@@href[1]{\endgroup#1\@@endlink}%
\providecommand \@sanitize@url [0]{\catcode `\\12\catcode `\$12\catcode
  `\&12\catcode `\#12\catcode `\^12\catcode `\_12\catcode `\%12\relax}%
\providecommand \@@startlink[1]{}%
\providecommand \@@endlink[0]{}%
\providecommand \url  [0]{\begingroup\@sanitize@url \@url }%
\providecommand \@url [1]{\endgroup\@href {#1}{\urlprefix }}%
\providecommand \urlprefix  [0]{URL }%
\providecommand \Eprint [0]{\href }%
\providecommand \doibase [0]{https://doi.org/}%
\providecommand \selectlanguage [0]{\@gobble}%
\providecommand \bibinfo  [0]{\@secondoftwo}%
\providecommand \bibfield  [0]{\@secondoftwo}%
\providecommand \translation [1]{[#1]}%
\providecommand \BibitemOpen [0]{}%
\providecommand \bibitemStop [0]{}%
\providecommand \bibitemNoStop [0]{.\EOS\space}%
\providecommand \EOS [0]{\spacefactor3000\relax}%
\providecommand \BibitemShut  [1]{\csname bibitem#1\endcsname}%
\let\auto@bib@innerbib\@empty
\bibitem [{\citenamefont {Atiyah}(1990)}]{atiyah_1990}%
  \BibitemOpen
  \bibfield  {author} {\bibinfo {author} {\bibfnamefont {M.}~\bibnamefont
  {Atiyah}},\ }\href {https://doi.org/10.1017/CBO9780511623868} {\emph
  {\bibinfo {title} {The Geometry and Physics of Knots}}},\ Lezioni Lincee\
  (\bibinfo  {publisher} {Cambridge University Press},\ \bibinfo {year}
  {1990})\BibitemShut {NoStop}%
\bibitem [{\citenamefont {Wu}\ \emph {et~al.}(2019)\citenamefont {Wu},
  \citenamefont {Soluyanov},\ and\ \citenamefont {Tomáš}}]{RN4}%
  \BibitemOpen
  \bibfield  {author} {\bibinfo {author} {\bibfnamefont {Q.}~\bibnamefont
  {Wu}}, \bibinfo {author} {\bibfnamefont {A.~A.}\ \bibnamefont {Soluyanov}},\
  and\ \bibinfo {author} {\bibfnamefont {B.}~\bibnamefont {Tomáš}},\
  }\bibfield  {title} {\bibinfo {title} {Non-abelian band topology in
  noninteracting metals},\ }\href {https://doi.org/doi:10.1126/science.aau8740}
  {\bibfield  {journal} {\bibinfo  {journal} {Science}\ }\textbf {\bibinfo
  {volume} {365}},\ \bibinfo {pages} {1273} (\bibinfo {year}
  {2019})}\BibitemShut {NoStop}%
\bibitem [{\citenamefont {Lee}\ \emph {et~al.}(2020)\citenamefont {Lee},
  \citenamefont {Sutrisno}, \citenamefont {Hofmann}, \citenamefont {Helbig},
  \citenamefont {Liu}, \citenamefont {Ang}, \citenamefont {Ang}, \citenamefont
  {Zhang}, \citenamefont {Greiter},\ and\ \citenamefont {Thomale}}]{RN9}%
  \BibitemOpen
  \bibfield  {author} {\bibinfo {author} {\bibfnamefont {C.~H.}\ \bibnamefont
  {Lee}}, \bibinfo {author} {\bibfnamefont {A.}~\bibnamefont {Sutrisno}},
  \bibinfo {author} {\bibfnamefont {T.}~\bibnamefont {Hofmann}}, \bibinfo
  {author} {\bibfnamefont {T.}~\bibnamefont {Helbig}}, \bibinfo {author}
  {\bibfnamefont {Y.}~\bibnamefont {Liu}}, \bibinfo {author} {\bibfnamefont
  {Y.~S.}\ \bibnamefont {Ang}}, \bibinfo {author} {\bibfnamefont {L.~K.}\
  \bibnamefont {Ang}}, \bibinfo {author} {\bibfnamefont {X.}~\bibnamefont
  {Zhang}}, \bibinfo {author} {\bibfnamefont {M.}~\bibnamefont {Greiter}},\
  and\ \bibinfo {author} {\bibfnamefont {R.}~\bibnamefont {Thomale}},\
  }\bibfield  {title} {\bibinfo {title} {Imaging nodal knots in momentum space
  through topolectrical circuits},\ }\href
  {https://doi.org/10.1038/s41467-020-17716-1} {\bibfield  {journal} {\bibinfo
  {journal} {Nature Communications}\ }\textbf {\bibinfo {volume} {11}},\
  \bibinfo {pages} {4385} (\bibinfo {year} {2020})}\BibitemShut {NoStop}%
\bibitem [{\citenamefont {Qiu}\ \emph {et~al.}(2023)\citenamefont {Qiu},
  \citenamefont {Zhang}, \citenamefont {Liu}, \citenamefont {Fan},
  \citenamefont {Zhang},\ and\ \citenamefont {Qiu}}]{RN14}%
  \BibitemOpen
  \bibfield  {author} {\bibinfo {author} {\bibfnamefont {H.}~\bibnamefont
  {Qiu}}, \bibinfo {author} {\bibfnamefont {Q.}~\bibnamefont {Zhang}}, \bibinfo
  {author} {\bibfnamefont {T.}~\bibnamefont {Liu}}, \bibinfo {author}
  {\bibfnamefont {X.}~\bibnamefont {Fan}}, \bibinfo {author} {\bibfnamefont
  {F.}~\bibnamefont {Zhang}},\ and\ \bibinfo {author} {\bibfnamefont
  {C.}~\bibnamefont {Qiu}},\ }\bibfield  {title} {\bibinfo {title} {Minimal
  non-abelian nodal braiding in ideal metamaterials},\ }\href
  {https://doi.org/10.1038/s41467-023-36952-9} {\bibfield  {journal} {\bibinfo
  {journal} {Nature Communications}\ }\textbf {\bibinfo {volume} {14}},\
  \bibinfo {pages} {1261} (\bibinfo {year} {2023})}\BibitemShut {NoStop}%
\bibitem [{\citenamefont {Leach}\ \emph {et~al.}(2004)\citenamefont {Leach},
  \citenamefont {Dennis}, \citenamefont {Courtial},\ and\ \citenamefont
  {Padgett}}]{RN2}%
  \BibitemOpen
  \bibfield  {author} {\bibinfo {author} {\bibfnamefont {J.}~\bibnamefont
  {Leach}}, \bibinfo {author} {\bibfnamefont {M.~R.}\ \bibnamefont {Dennis}},
  \bibinfo {author} {\bibfnamefont {J.}~\bibnamefont {Courtial}},\ and\
  \bibinfo {author} {\bibfnamefont {M.~J.}\ \bibnamefont {Padgett}},\
  }\bibfield  {title} {\bibinfo {title} {Knotted threads of darkness},\ }\href
  {https://doi.org/10.1038/432165a} {\bibfield  {journal} {\bibinfo  {journal}
  {Nature}\ }\textbf {\bibinfo {volume} {432}},\ \bibinfo {pages} {165}
  (\bibinfo {year} {2004})}\BibitemShut {NoStop}%
\bibitem [{\citenamefont {Pisanty}\ \emph {et~al.}(2019)\citenamefont
  {Pisanty}, \citenamefont {Machado}, \citenamefont {Vicuña-Hernández},
  \citenamefont {Picón}, \citenamefont {Celi}, \citenamefont {Torres},\ and\
  \citenamefont {Lewenstein}}]{RN6}%
  \BibitemOpen
  \bibfield  {author} {\bibinfo {author} {\bibfnamefont {E.}~\bibnamefont
  {Pisanty}}, \bibinfo {author} {\bibfnamefont {G.~J.}\ \bibnamefont
  {Machado}}, \bibinfo {author} {\bibfnamefont {V.}~\bibnamefont
  {Vicuña-Hernández}}, \bibinfo {author} {\bibfnamefont {A.}~\bibnamefont
  {Picón}}, \bibinfo {author} {\bibfnamefont {A.}~\bibnamefont {Celi}},
  \bibinfo {author} {\bibfnamefont {J.~P.}\ \bibnamefont {Torres}},\ and\
  \bibinfo {author} {\bibfnamefont {M.}~\bibnamefont {Lewenstein}},\ }\bibfield
   {title} {\bibinfo {title} {Knotting fractional-order knots with the
  polarization state of light},\ }\href
  {https://doi.org/10.1038/s41566-019-0450-2} {\bibfield  {journal} {\bibinfo
  {journal} {Nature Photonics}\ }\textbf {\bibinfo {volume} {13}},\ \bibinfo
  {pages} {569} (\bibinfo {year} {2019})}\BibitemShut {NoStop}%
\bibitem [{\citenamefont {Yang}\ \emph {et~al.}(2020)\citenamefont {Yang},
  \citenamefont {Chiu}, \citenamefont {Fang},\ and\ \citenamefont
  {Hu}}]{Jones_Polynomial_Yang}%
  \BibitemOpen
  \bibfield  {author} {\bibinfo {author} {\bibfnamefont {Z.}~\bibnamefont
  {Yang}}, \bibinfo {author} {\bibfnamefont {C.-K.}\ \bibnamefont {Chiu}},
  \bibinfo {author} {\bibfnamefont {C.}~\bibnamefont {Fang}},\ and\ \bibinfo
  {author} {\bibfnamefont {J.}~\bibnamefont {Hu}},\ }\bibfield  {title}
  {\bibinfo {title} {Jones polynomial and knot transitions in hermitian and
  non-hermitian topological semimetals},\ }\href
  {https://doi.org/10.1103/PhysRevLett.124.186402} {\bibfield  {journal}
  {\bibinfo  {journal} {Physical Review Letters}\ }\textbf {\bibinfo {volume}
  {124}},\ \bibinfo {pages} {186402} (\bibinfo {year} {2020})}\BibitemShut
  {NoStop}%
\bibitem [{\citenamefont {Wang}\ \emph {et~al.}(2017)\citenamefont {Wang},
  \citenamefont {Zhang}, \citenamefont {Chen}, \citenamefont {Yu},\ and\
  \citenamefont {Zhai}}]{wang2017scheme}%
  \BibitemOpen
  \bibfield  {author} {\bibinfo {author} {\bibfnamefont {C.}~\bibnamefont
  {Wang}}, \bibinfo {author} {\bibfnamefont {P.}~\bibnamefont {Zhang}},
  \bibinfo {author} {\bibfnamefont {X.}~\bibnamefont {Chen}}, \bibinfo {author}
  {\bibfnamefont {J.}~\bibnamefont {Yu}},\ and\ \bibinfo {author}
  {\bibfnamefont {H.}~\bibnamefont {Zhai}},\ }\bibfield  {title} {\bibinfo
  {title} {Scheme to measure the topological number of a chern insulator from
  quench dynamics},\ }\href {https://doi.org/10.1103/PhysRevLett.118.185701}
  {\bibfield  {journal} {\bibinfo  {journal} {Physical Review Letters}\
  }\textbf {\bibinfo {volume} {118}},\ \bibinfo {pages} {185701} (\bibinfo
  {year} {2017})}\BibitemShut {NoStop}%
\bibitem [{\citenamefont {Yang}(1967)}]{YangCN_yangbaxter}%
  \BibitemOpen
  \bibfield  {author} {\bibinfo {author} {\bibfnamefont {C.~N.}\ \bibnamefont
  {Yang}},\ }\bibfield  {title} {\bibinfo {title} {Some exact results for the
  many-body problem in one dimension with repulsive delta-function
  interaction},\ }\href {https://doi.org/10.1103/PhysRevLett.19.1312}
  {\bibfield  {journal} {\bibinfo  {journal} {Physical Review Letters}\
  }\textbf {\bibinfo {volume} {19}},\ \bibinfo {pages} {1312} (\bibinfo {year}
  {1967})}\BibitemShut {NoStop}%
\bibitem [{\citenamefont {JIMBO}(1989)}]{Yang_baxter}%
  \BibitemOpen
  \bibfield  {author} {\bibinfo {author} {\bibfnamefont {M.}~\bibnamefont
  {JIMBO}},\ }\bibfield  {title} {\bibinfo {title} {Introduction to the
  yang-baxter equation},\ }\href {https://doi.org/10.1142/S0217751X89001503}
  {\bibfield  {journal} {\bibinfo  {journal} {International Journal of Modern
  Physics A}\ }\textbf {\bibinfo {volume} {04}},\ \bibinfo {pages} {3759}
  (\bibinfo {year} {1989})},\ \Eprint
  {https://arxiv.org/abs/https://doi.org/10.1142/S0217751X89001503}
  {https://doi.org/10.1142/S0217751X89001503} \BibitemShut {NoStop}%
\bibitem [{\citenamefont {Liu}\ \emph {et~al.}(2022)\citenamefont {Liu},
  \citenamefont {Gao}, \citenamefont {Wang}, \citenamefont {Xi}, \citenamefont
  {Hu}, \citenamefont {Wang}, \citenamefont {Liu}, \citenamefont {Lin},
  \citenamefont {Deng}, \citenamefont {Yang}, \citenamefont {Zhou},
  \citenamefont {Yang}, \citenamefont {Chong},\ and\ \citenamefont
  {Zhang}}]{Topological_Chern_vectors}%
  \BibitemOpen
  \bibfield  {author} {\bibinfo {author} {\bibfnamefont {G.-G.}\ \bibnamefont
  {Liu}}, \bibinfo {author} {\bibfnamefont {Z.}~\bibnamefont {Gao}}, \bibinfo
  {author} {\bibfnamefont {Q.}~\bibnamefont {Wang}}, \bibinfo {author}
  {\bibfnamefont {X.}~\bibnamefont {Xi}}, \bibinfo {author} {\bibfnamefont
  {Y.-H.}\ \bibnamefont {Hu}}, \bibinfo {author} {\bibfnamefont
  {M.}~\bibnamefont {Wang}}, \bibinfo {author} {\bibfnamefont {C.}~\bibnamefont
  {Liu}}, \bibinfo {author} {\bibfnamefont {X.}~\bibnamefont {Lin}}, \bibinfo
  {author} {\bibfnamefont {L.}~\bibnamefont {Deng}}, \bibinfo {author}
  {\bibfnamefont {S.~A.}\ \bibnamefont {Yang}}, \bibinfo {author}
  {\bibfnamefont {P.}~\bibnamefont {Zhou}}, \bibinfo {author} {\bibfnamefont
  {Y.}~\bibnamefont {Yang}}, \bibinfo {author} {\bibfnamefont {Y.}~\bibnamefont
  {Chong}},\ and\ \bibinfo {author} {\bibfnamefont {B.}~\bibnamefont {Zhang}},\
  }\bibfield  {title} {\bibinfo {title} {Topological chern vectors in
  three-dimensional photonic crystals},\ }\href
  {https://doi.org/10.1038/s41586-022-05077-2} {\bibfield  {journal} {\bibinfo
  {journal} {Nature}\ }\textbf {\bibinfo {volume} {609}},\ \bibinfo {pages}
  {925–930} (\bibinfo {year} {2022})}\BibitemShut {NoStop}%
\bibitem [{\citenamefont {Sun}\ \emph {et~al.}(2017)\citenamefont {Sun},
  \citenamefont {Lian},\ and\ \citenamefont
  {Zhang}}]{Double_Helix_Superconductor}%
  \BibitemOpen
  \bibfield  {author} {\bibinfo {author} {\bibfnamefont {X.-Q.}\ \bibnamefont
  {Sun}}, \bibinfo {author} {\bibfnamefont {B.}~\bibnamefont {Lian}},\ and\
  \bibinfo {author} {\bibfnamefont {S.-C.}\ \bibnamefont {Zhang}},\ }\bibfield
  {title} {\bibinfo {title} {Double helix nodal line superconductor},\ }\href
  {https://doi.org/10.1103/PhysRevLett.119.147001} {\bibfield  {journal}
  {\bibinfo  {journal} {Physical Review Letters}\ }\textbf {\bibinfo {volume}
  {119}},\ \bibinfo {pages} {147001} (\bibinfo {year} {2017})}\BibitemShut
  {NoStop}%
\bibitem [{\citenamefont {Ren}\ and\ \citenamefont
  {Sinitsyn}(2013)}]{PhysRevE.87.050101}%
  \BibitemOpen
  \bibfield  {author} {\bibinfo {author} {\bibfnamefont {J.}~\bibnamefont
  {Ren}}\ and\ \bibinfo {author} {\bibfnamefont {N.~A.}\ \bibnamefont
  {Sinitsyn}},\ }\bibfield  {title} {\bibinfo {title} {Braid group and
  topological phase transitions in nonequilibrium stochastic dynamics},\ }\href
  {https://doi.org/10.1103/PhysRevE.87.050101} {\bibfield  {journal} {\bibinfo
  {journal} {Physical Review E}\ }\textbf {\bibinfo {volume} {87}},\ \bibinfo
  {pages} {050101(R)} (\bibinfo {year} {2013})}\BibitemShut {NoStop}%
\bibitem [{\citenamefont {Engelhardt}\ \emph {et~al.}(2017)\citenamefont
  {Engelhardt}, \citenamefont {Benito}, \citenamefont {Platero}, \citenamefont
  {Schaller},\ and\ \citenamefont {Brandes}}]{PhysRevB.96.241404}%
  \BibitemOpen
  \bibfield  {author} {\bibinfo {author} {\bibfnamefont {G.}~\bibnamefont
  {Engelhardt}}, \bibinfo {author} {\bibfnamefont {M.}~\bibnamefont {Benito}},
  \bibinfo {author} {\bibfnamefont {G.}~\bibnamefont {Platero}}, \bibinfo
  {author} {\bibfnamefont {G.}~\bibnamefont {Schaller}},\ and\ \bibinfo
  {author} {\bibfnamefont {T.}~\bibnamefont {Brandes}},\ }\bibfield  {title}
  {\bibinfo {title} {Random-walk topological transition revealed via electron
  counting},\ }\href {https://doi.org/10.1103/PhysRevB.96.241404} {\bibfield
  {journal} {\bibinfo  {journal} {Physical Review B}\ }\textbf {\bibinfo
  {volume} {96}},\ \bibinfo {pages} {241404(R)} (\bibinfo {year}
  {2017})}\BibitemShut {NoStop}%
\bibitem [{\citenamefont {Wojcik}\ \emph {et~al.}(2020)\citenamefont {Wojcik},
  \citenamefont {Sun}, \citenamefont {Bzdu\ifmmode~\check{s}\else
  \v{s}\fi{}ek},\ and\ \citenamefont {Fan}}]{PhysRevB.101.205417}%
  \BibitemOpen
  \bibfield  {author} {\bibinfo {author} {\bibfnamefont {C.~C.}\ \bibnamefont
  {Wojcik}}, \bibinfo {author} {\bibfnamefont {X.-Q.}\ \bibnamefont {Sun}},
  \bibinfo {author} {\bibfnamefont {T.~c.~v.}\ \bibnamefont
  {Bzdu\ifmmode~\check{s}\else \v{s}\fi{}ek}},\ and\ \bibinfo {author}
  {\bibfnamefont {S.}~\bibnamefont {Fan}},\ }\bibfield  {title} {\bibinfo
  {title} {Homotopy characterization of non-hermitian hamiltonians},\ }\href
  {https://doi.org/10.1103/PhysRevB.101.205417} {\bibfield  {journal} {\bibinfo
   {journal} {Physical Review B}\ }\textbf {\bibinfo {volume} {101}},\ \bibinfo
  {pages} {205417} (\bibinfo {year} {2020})}\BibitemShut {NoStop}%
\bibitem [{\citenamefont {Hu}\ and\ \citenamefont
  {Zhao}(2021)}]{PhysRevLett.126.010401}%
  \BibitemOpen
  \bibfield  {author} {\bibinfo {author} {\bibfnamefont {H.}~\bibnamefont
  {Hu}}\ and\ \bibinfo {author} {\bibfnamefont {E.}~\bibnamefont {Zhao}},\
  }\bibfield  {title} {\bibinfo {title} {Knots and non-hermitian bloch bands},\
  }\href {https://doi.org/10.1103/PhysRevLett.126.010401} {\bibfield  {journal}
  {\bibinfo  {journal} {Physical Review Letters}\ }\textbf {\bibinfo {volume}
  {126}},\ \bibinfo {pages} {010401} (\bibinfo {year} {2021})}\BibitemShut
  {NoStop}%
\bibitem [{\citenamefont {Wang}\ \emph
  {et~al.}(2021{\natexlab{a}})\citenamefont {Wang}, \citenamefont {Dutt},
  \citenamefont {Yang}, \citenamefont {Wojcik}, \citenamefont {VUČKOVIĆ},\
  and\ \citenamefont {Fan}}]{RN13}%
  \BibitemOpen
  \bibfield  {author} {\bibinfo {author} {\bibfnamefont {K.}~\bibnamefont
  {Wang}}, \bibinfo {author} {\bibfnamefont {A.}~\bibnamefont {Dutt}}, \bibinfo
  {author} {\bibfnamefont {K.~Y.}\ \bibnamefont {Yang}}, \bibinfo {author}
  {\bibfnamefont {C.~C.}\ \bibnamefont {Wojcik}}, \bibinfo {author}
  {\bibfnamefont {J.}~\bibnamefont {VUČKOVIĆ}},\ and\ \bibinfo {author}
  {\bibfnamefont {S.}~\bibnamefont {Fan}},\ }\bibfield  {title} {\bibinfo
  {title} {Generating arbitrary topological windings of a non-hermitian band},\
  }\href {https://doi.org/doi:10.1126/science.abf6568} {\bibfield  {journal}
  {\bibinfo  {journal} {Science}\ }\textbf {\bibinfo {volume} {371}},\ \bibinfo
  {pages} {1240} (\bibinfo {year} {2021}{\natexlab{a}})}\BibitemShut {NoStop}%
\bibitem [{\citenamefont {Wang}\ \emph
  {et~al.}(2021{\natexlab{b}})\citenamefont {Wang}, \citenamefont {Dutt},
  \citenamefont {Wojcik},\ and\ \citenamefont {Fan}}]{RN5}%
  \BibitemOpen
  \bibfield  {author} {\bibinfo {author} {\bibfnamefont {K.}~\bibnamefont
  {Wang}}, \bibinfo {author} {\bibfnamefont {A.}~\bibnamefont {Dutt}}, \bibinfo
  {author} {\bibfnamefont {C.~C.}\ \bibnamefont {Wojcik}},\ and\ \bibinfo
  {author} {\bibfnamefont {S.}~\bibnamefont {Fan}},\ }\bibfield  {title}
  {\bibinfo {title} {Topological complex-energy braiding of non-hermitian
  bands},\ }\href {https://doi.org/10.1038/s41586-021-03848-x} {\bibfield
  {journal} {\bibinfo  {journal} {Nature}\ }\textbf {\bibinfo {volume} {598}},\
  \bibinfo {pages} {59} (\bibinfo {year} {2021}{\natexlab{b}})}\BibitemShut
  {NoStop}%
\bibitem [{\citenamefont {Patil}\ \emph {et~al.}(2022)\citenamefont {Patil},
  \citenamefont {Höller}, \citenamefont {Henry}, \citenamefont {Guria},
  \citenamefont {Zhang}, \citenamefont {Jiang}, \citenamefont {Kralj},
  \citenamefont {Read},\ and\ \citenamefont {Harris}}]{Patil_nature_2022}%
  \BibitemOpen
  \bibfield  {author} {\bibinfo {author} {\bibfnamefont {Y.~S.~S.}\
  \bibnamefont {Patil}}, \bibinfo {author} {\bibfnamefont {J.}~\bibnamefont
  {Höller}}, \bibinfo {author} {\bibfnamefont {P.~A.}\ \bibnamefont {Henry}},
  \bibinfo {author} {\bibfnamefont {C.}~\bibnamefont {Guria}}, \bibinfo
  {author} {\bibfnamefont {Y.}~\bibnamefont {Zhang}}, \bibinfo {author}
  {\bibfnamefont {L.}~\bibnamefont {Jiang}}, \bibinfo {author} {\bibfnamefont
  {N.}~\bibnamefont {Kralj}}, \bibinfo {author} {\bibfnamefont
  {N.}~\bibnamefont {Read}},\ and\ \bibinfo {author} {\bibfnamefont {J.~G.~E.}\
  \bibnamefont {Harris}},\ }\bibfield  {title} {\bibinfo {title} {Measuring the
  knot of non-hermitian degeneracies and non-commuting braids},\ }\href
  {https://doi.org/10.1038/s41586-022-04796-w} {\bibfield  {journal} {\bibinfo
  {journal} {Nature}\ }\textbf {\bibinfo {volume} {607}},\ \bibinfo {pages}
  {271} (\bibinfo {year} {2022})}\BibitemShut {NoStop}%
\bibitem [{\citenamefont {Yu}\ \emph {et~al.}(2022)\citenamefont {Yu},
  \citenamefont {Yu}, \citenamefont {Zhang}, \citenamefont {Zhang},
  \citenamefont {Ouyang}, \citenamefont {Liu}, \citenamefont {Deng},\ and\
  \citenamefont {Duan}}]{DDL_npj_2022}%
  \BibitemOpen
  \bibfield  {author} {\bibinfo {author} {\bibfnamefont {Y.}~\bibnamefont
  {Yu}}, \bibinfo {author} {\bibfnamefont {L.-W.}\ \bibnamefont {Yu}}, \bibinfo
  {author} {\bibfnamefont {W.}~\bibnamefont {Zhang}}, \bibinfo {author}
  {\bibfnamefont {H.}~\bibnamefont {Zhang}}, \bibinfo {author} {\bibfnamefont
  {X.}~\bibnamefont {Ouyang}}, \bibinfo {author} {\bibfnamefont
  {Y.}~\bibnamefont {Liu}}, \bibinfo {author} {\bibfnamefont {D.-L.}\
  \bibnamefont {Deng}},\ and\ \bibinfo {author} {\bibfnamefont {L.~M.}\
  \bibnamefont {Duan}},\ }\bibfield  {title} {\bibinfo {title} {Experimental
  unsupervised learning of non-hermitian knotted phases with solid-state
  spins},\ }\href {https://doi.org/10.1038/s41534-022-00629-w} {\bibfield
  {journal} {\bibinfo  {journal} {npj Quantum Inform.}\ }\textbf {\bibinfo
  {volume} {8}},\ \bibinfo {pages} {116} (\bibinfo {year} {2022})}\BibitemShut
  {NoStop}%
\bibitem [{\citenamefont {Midya}(2023)}]{Midya_APL_2023}%
  \BibitemOpen
  \bibfield  {author} {\bibinfo {author} {\bibfnamefont {B.}~\bibnamefont
  {Midya}},\ }\bibfield  {title} {\bibinfo {title} {Gain-loss-induced
  non-abelian bloch braids},\ }\href {https://doi.org/10.1063/5.0164009}
  {\bibfield  {journal} {\bibinfo  {journal} {Applied Physics Letters}\
  }\textbf {\bibinfo {volume} {123}},\ \bibinfo {pages} {123101} (\bibinfo
  {year} {2023})}\BibitemShut {NoStop}%
\bibitem [{\citenamefont {Li}\ \emph {et~al.}(2021)\citenamefont {Li},
  \citenamefont {Mu}, \citenamefont {Lee},\ and\ \citenamefont
  {Gong}}]{GongJiangbin_nc_2021}%
  \BibitemOpen
  \bibfield  {author} {\bibinfo {author} {\bibfnamefont {L.}~\bibnamefont
  {Li}}, \bibinfo {author} {\bibfnamefont {S.}~\bibnamefont {Mu}}, \bibinfo
  {author} {\bibfnamefont {C.~H.}\ \bibnamefont {Lee}},\ and\ \bibinfo {author}
  {\bibfnamefont {J.}~\bibnamefont {Gong}},\ }\bibfield  {title} {\bibinfo
  {title} {Quantized classical response from spectral winding topology},\
  }\href {https://doi.org/10.1038/s41467-021-25626-z} {\bibfield  {journal}
  {\bibinfo  {journal} {Nature Communications}\ }\textbf {\bibinfo {volume}
  {12}},\ \bibinfo {pages} {5294} (\bibinfo {year} {2021})}\BibitemShut
  {NoStop}%
\bibitem [{\citenamefont {Ozsváth}\ and\ \citenamefont
  {Szabó}(2004)}]{Peter_AM_2004}%
  \BibitemOpen
  \bibfield  {author} {\bibinfo {author} {\bibfnamefont {P.}~\bibnamefont
  {Ozsváth}}\ and\ \bibinfo {author} {\bibfnamefont {Z.}~\bibnamefont
  {Szabó}},\ }\bibfield  {title} {\bibinfo {title} {Holomorphic disks and knot
  invariants},\ }\href
  {https://doi.org/https://doi.org/10.1016/j.aim.2003.05.001} {\bibfield
  {journal} {\bibinfo  {journal} {Advances in Mathematics}\ }\textbf {\bibinfo
  {volume} {186}},\ \bibinfo {pages} {58} (\bibinfo {year} {2004})}\BibitemShut
  {NoStop}%
\bibitem [{\citenamefont {Witten}(1989)}]{Jones_polynomial}%
  \BibitemOpen
  \bibfield  {author} {\bibinfo {author} {\bibfnamefont {E.}~\bibnamefont
  {Witten}},\ }\bibfield  {title} {\bibinfo {title} {Quantum field theory and
  the jones polynomial},\ }\href {https://doi.org/10.1007/BF01217730}
  {\bibfield  {journal} {\bibinfo  {journal} {Communications in Mathematical
  Physics}\ }\textbf {\bibinfo {volume} {121}},\ \bibinfo {pages} {351}
  (\bibinfo {year} {1989})}\BibitemShut {NoStop}%
\bibitem [{\citenamefont {Vandans}\ \emph {et~al.}(2020)\citenamefont
  {Vandans}, \citenamefont {Yang}, \citenamefont {Wu},\ and\ \citenamefont
  {Dai}}]{Vandans_pre_2020}%
  \BibitemOpen
  \bibfield  {author} {\bibinfo {author} {\bibfnamefont {O.}~\bibnamefont
  {Vandans}}, \bibinfo {author} {\bibfnamefont {K.}~\bibnamefont {Yang}},
  \bibinfo {author} {\bibfnamefont {Z.}~\bibnamefont {Wu}},\ and\ \bibinfo
  {author} {\bibfnamefont {L.}~\bibnamefont {Dai}},\ }\bibfield  {title}
  {\bibinfo {title} {Identifying knot types of polymer conformations by machine
  learning},\ }\href {https://doi.org/10.1103/PhysRevE.101.022502} {\bibfield
  {journal} {\bibinfo  {journal} {Physical Review E}\ }\textbf {\bibinfo
  {volume} {101}},\ \bibinfo {pages} {022502} (\bibinfo {year}
  {2020})}\BibitemShut {NoStop}%
\bibitem [{\citenamefont {Davies}\ \emph {et~al.}(2021)\citenamefont {Davies},
  \citenamefont {Velickovic}, \citenamefont {Buesing}, \citenamefont
  {Blackwell}, \citenamefont {Zheng}, \citenamefont {Tomasev}, \citenamefont
  {Tanburn}, \citenamefont {Battaglia}, \citenamefont {Blundell}, \citenamefont
  {Juhasz}, \citenamefont {Lackenby}, \citenamefont {Williamson}, \citenamefont
  {Hassabis},\ and\ \citenamefont {Kohli}}]{DaviesAlex_natrue_2021}%
  \BibitemOpen
  \bibfield  {author} {\bibinfo {author} {\bibfnamefont {A.}~\bibnamefont
  {Davies}}, \bibinfo {author} {\bibfnamefont {P.}~\bibnamefont {Velickovic}},
  \bibinfo {author} {\bibfnamefont {L.}~\bibnamefont {Buesing}}, \bibinfo
  {author} {\bibfnamefont {S.}~\bibnamefont {Blackwell}}, \bibinfo {author}
  {\bibfnamefont {D.}~\bibnamefont {Zheng}}, \bibinfo {author} {\bibfnamefont
  {N.}~\bibnamefont {Tomasev}}, \bibinfo {author} {\bibfnamefont
  {R.}~\bibnamefont {Tanburn}}, \bibinfo {author} {\bibfnamefont
  {P.}~\bibnamefont {Battaglia}}, \bibinfo {author} {\bibfnamefont
  {C.}~\bibnamefont {Blundell}}, \bibinfo {author} {\bibfnamefont
  {A.}~\bibnamefont {Juhasz}}, \bibinfo {author} {\bibfnamefont
  {M.}~\bibnamefont {Lackenby}}, \bibinfo {author} {\bibfnamefont
  {G.}~\bibnamefont {Williamson}}, \bibinfo {author} {\bibfnamefont
  {D.}~\bibnamefont {Hassabis}},\ and\ \bibinfo {author} {\bibfnamefont
  {P.}~\bibnamefont {Kohli}},\ }\bibfield  {title} {\bibinfo {title} {Advancing
  mathematics by guiding human intuition with ai},\ }\href
  {https://doi.org/10.1038/s41586-021-04086-x} {\bibfield  {journal} {\bibinfo
  {journal} {Nature}\ }\textbf {\bibinfo {volume} {600}},\ \bibinfo {pages}
  {70} (\bibinfo {year} {2021})}\BibitemShut {NoStop}%
\bibitem [{\citenamefont {Wang}(2016)}]{WangLei_prb_2016}%
  \BibitemOpen
  \bibfield  {author} {\bibinfo {author} {\bibfnamefont {L.}~\bibnamefont
  {Wang}},\ }\bibfield  {title} {\bibinfo {title} {Discovering phase
  transitions with unsupervised learning},\ }\href
  {https://doi.org/10.1103/PhysRevB.94.195105} {\bibfield  {journal} {\bibinfo
  {journal} {Physical Review B}\ }\textbf {\bibinfo {volume} {94}},\ \bibinfo
  {pages} {195105} (\bibinfo {year} {2016})}\BibitemShut {NoStop}%
\bibitem [{\citenamefont {Wang}\ and\ \citenamefont
  {Zhai}(2017)}]{wang2017machine}%
  \BibitemOpen
  \bibfield  {author} {\bibinfo {author} {\bibfnamefont {C.}~\bibnamefont
  {Wang}}\ and\ \bibinfo {author} {\bibfnamefont {H.}~\bibnamefont {Zhai}},\
  }\bibfield  {title} {\bibinfo {title} {Machine learning of frustrated
  classical spin models. i. principal component analysis},\ }\href
  {https://doi.org/10.1103/PhysRevB.96.144432} {\bibfield  {journal} {\bibinfo
  {journal} {Phys. Rev. B}\ }\textbf {\bibinfo {volume} {96}},\ \bibinfo
  {pages} {144432} (\bibinfo {year} {2017})}\BibitemShut {NoStop}%
\bibitem [{\citenamefont {Wetzel}(2017)}]{Wetzel_pre_2017}%
  \BibitemOpen
  \bibfield  {author} {\bibinfo {author} {\bibfnamefont {S.~J.}\ \bibnamefont
  {Wetzel}},\ }\bibfield  {title} {\bibinfo {title} {Unsupervised learning of
  phase transitions: From principal component analysis to variational
  autoencoders},\ }\href {https://doi.org/10.1103/PhysRevE.96.022140}
  {\bibfield  {journal} {\bibinfo  {journal} {Physical Review E}\ }\textbf
  {\bibinfo {volume} {96}},\ \bibinfo {pages} {022140} (\bibinfo {year}
  {2017})}\BibitemShut {NoStop}%
\bibitem [{\citenamefont {Rodriguez-Nieva}\ and\ \citenamefont
  {Scheurer}(2019)}]{Rodriguez-Nieva_Natphys_2019}%
  \BibitemOpen
  \bibfield  {author} {\bibinfo {author} {\bibfnamefont {J.~F.}\ \bibnamefont
  {Rodriguez-Nieva}}\ and\ \bibinfo {author} {\bibfnamefont {M.~S.}\
  \bibnamefont {Scheurer}},\ }\bibfield  {title} {\bibinfo {title} {Identifying
  topological order through unsupervised machine learning},\ }\href
  {https://doi.org/10.1038/s41567-019-0512-x} {\bibfield  {journal} {\bibinfo
  {journal} {Nature Physics}\ }\textbf {\bibinfo {volume} {15}},\ \bibinfo
  {pages} {790} (\bibinfo {year} {2019})}\BibitemShut {NoStop}%
\bibitem [{\citenamefont {Lidiak}\ and\ \citenamefont
  {Gong}(2020)}]{Lidiak_prl_2020}%
  \BibitemOpen
  \bibfield  {author} {\bibinfo {author} {\bibfnamefont {A.}~\bibnamefont
  {Lidiak}}\ and\ \bibinfo {author} {\bibfnamefont {Z.}~\bibnamefont {Gong}},\
  }\bibfield  {title} {\bibinfo {title} {Unsupervised machine learning of
  quantum phase transitions using diffusion maps},\ }\href
  {https://doi.org/10.1103/PhysRevLett.125.225701} {\bibfield  {journal}
  {\bibinfo  {journal} {Physical Review Letters}\ }\textbf {\bibinfo {volume}
  {125}},\ \bibinfo {pages} {225701} (\bibinfo {year} {2020})}\BibitemShut
  {NoStop}%
\bibitem [{\citenamefont {Long}\ \emph {et~al.}(2020)\citenamefont {Long},
  \citenamefont {Ren},\ and\ \citenamefont {Chen}}]{LongYang_prl_2020}%
  \BibitemOpen
  \bibfield  {author} {\bibinfo {author} {\bibfnamefont {Y.}~\bibnamefont
  {Long}}, \bibinfo {author} {\bibfnamefont {J.}~\bibnamefont {Ren}},\ and\
  \bibinfo {author} {\bibfnamefont {H.}~\bibnamefont {Chen}},\ }\bibfield
  {title} {\bibinfo {title} {Unsupervised manifold clustering of topological
  phononics},\ }\href {https://doi.org/10.1103/PhysRevLett.124.185501}
  {\bibfield  {journal} {\bibinfo  {journal} {Physical Review Letters}\
  }\textbf {\bibinfo {volume} {124}},\ \bibinfo {pages} {185501} (\bibinfo
  {year} {2020})}\BibitemShut {NoStop}%
\bibitem [{\citenamefont {Scheurer}\ and\ \citenamefont
  {Slager}(2020)}]{Scheurer_prl_2020}%
  \BibitemOpen
  \bibfield  {author} {\bibinfo {author} {\bibfnamefont {M.~S.}\ \bibnamefont
  {Scheurer}}\ and\ \bibinfo {author} {\bibfnamefont {R.-J.}\ \bibnamefont
  {Slager}},\ }\bibfield  {title} {\bibinfo {title} {Unsupervised machine
  learning and band topology},\ }\href
  {https://doi.org/10.1103/PhysRevLett.124.226401} {\bibfield  {journal}
  {\bibinfo  {journal} {Physical Review Letters}\ }\textbf {\bibinfo {volume}
  {124}},\ \bibinfo {pages} {226401} (\bibinfo {year} {2020})}\BibitemShut
  {NoStop}%
\bibitem [{\citenamefont {Che}\ \emph {et~al.}(2020)\citenamefont {Che},
  \citenamefont {Gneiting}, \citenamefont {Liu},\ and\ \citenamefont
  {Nori}}]{CheYanming_prb_2020}%
  \BibitemOpen
  \bibfield  {author} {\bibinfo {author} {\bibfnamefont {Y.}~\bibnamefont
  {Che}}, \bibinfo {author} {\bibfnamefont {C.}~\bibnamefont {Gneiting}},
  \bibinfo {author} {\bibfnamefont {T.}~\bibnamefont {Liu}},\ and\ \bibinfo
  {author} {\bibfnamefont {F.}~\bibnamefont {Nori}},\ }\bibfield  {title}
  {\bibinfo {title} {Topological quantum phase transitions retrieved through
  unsupervised machine learning},\ }\href
  {https://doi.org/10.1103/PhysRevB.102.134213} {\bibfield  {journal} {\bibinfo
   {journal} {Physical Review B}\ }\textbf {\bibinfo {volume} {102}},\ \bibinfo
  {pages} {134213} (\bibinfo {year} {2020})}\BibitemShut {NoStop}%
\bibitem [{\citenamefont {Yu}\ and\ \citenamefont
  {Deng}(2021)}]{YuLiwei_prl_2021}%
  \BibitemOpen
  \bibfield  {author} {\bibinfo {author} {\bibfnamefont {L.-W.}\ \bibnamefont
  {Yu}}\ and\ \bibinfo {author} {\bibfnamefont {D.-L.}\ \bibnamefont {Deng}},\
  }\bibfield  {title} {\bibinfo {title} {Unsupervised learning of non-hermitian
  topological phases},\ }\href {https://doi.org/10.1103/PhysRevLett.126.240402}
  {\bibfield  {journal} {\bibinfo  {journal} {Physical Review Letters}\
  }\textbf {\bibinfo {volume} {126}},\ \bibinfo {pages} {240402} (\bibinfo
  {year} {2021})}\BibitemShut {NoStop}%
\bibitem [{\citenamefont {Park}\ \emph {et~al.}(2022)\citenamefont {Park},
  \citenamefont {Hwang},\ and\ \citenamefont {Yang}}]{ParkSungjoon_prb_2022}%
  \BibitemOpen
  \bibfield  {author} {\bibinfo {author} {\bibfnamefont {S.}~\bibnamefont
  {Park}}, \bibinfo {author} {\bibfnamefont {Y.}~\bibnamefont {Hwang}},\ and\
  \bibinfo {author} {\bibfnamefont {B.-J.}\ \bibnamefont {Yang}},\ }\bibfield
  {title} {\bibinfo {title} {Unsupervised learning of topological phase diagram
  using topological data analysis},\ }\href
  {https://doi.org/10.1103/PhysRevB.105.195115} {\bibfield  {journal} {\bibinfo
   {journal} {Physical Review B}\ }\textbf {\bibinfo {volume} {105}},\ \bibinfo
  {pages} {195115} (\bibinfo {year} {2022})}\BibitemShut {NoStop}%
\bibitem [{\citenamefont {Li}\ \emph {et~al.}(2023)\citenamefont {Li},
  \citenamefont {Ao}, \citenamefont {Hu}, \citenamefont {Lu}, \citenamefont
  {Chan},\ and\ \citenamefont {Gong}}]{GongQH_LPR_2023}%
  \BibitemOpen
  \bibfield  {author} {\bibinfo {author} {\bibfnamefont {Y.}~\bibnamefont
  {Li}}, \bibinfo {author} {\bibfnamefont {Y.}~\bibnamefont {Ao}}, \bibinfo
  {author} {\bibfnamefont {X.}~\bibnamefont {Hu}}, \bibinfo {author}
  {\bibfnamefont {C.}~\bibnamefont {Lu}}, \bibinfo {author} {\bibfnamefont
  {C.~T.}\ \bibnamefont {Chan}},\ and\ \bibinfo {author} {\bibfnamefont
  {Q.}~\bibnamefont {Gong}},\ }\bibfield  {title} {\bibinfo {title}
  {Unsupervised learning of non-hermitian photonic bulk topology},\ }\href
  {https://doi.org/https://doi.org/10.1002/lpor.202300481} {\bibfield
  {journal} {\bibinfo  {journal} {Laser Photonics Reviews}\ ,\ \bibinfo {pages}
  {2300481}} (\bibinfo {year} {2023})}\BibitemShut {NoStop}%
\bibitem [{\citenamefont {Long}\ and\ \citenamefont
  {Zhang}(2023)}]{LongYang_prl_2023}%
  \BibitemOpen
  \bibfield  {author} {\bibinfo {author} {\bibfnamefont {Y.}~\bibnamefont
  {Long}}\ and\ \bibinfo {author} {\bibfnamefont {B.}~\bibnamefont {Zhang}},\
  }\bibfield  {title} {\bibinfo {title} {Unsupervised data-driven
  classification of topological gapped systems with symmetries},\ }\href
  {https://doi.org/10.1103/PhysRevLett.130.036601} {\bibfield  {journal}
  {\bibinfo  {journal} {Physical Review Letters}\ }\textbf {\bibinfo {volume}
  {130}},\ \bibinfo {pages} {036601} (\bibinfo {year} {2023})}\BibitemShut
  {NoStop}%
\bibitem [{\citenamefont {Coifman}\ \emph {et~al.}(2005)\citenamefont
  {Coifman}, \citenamefont {Lafon}, \citenamefont {Lee}, \citenamefont
  {Maggioni}, \citenamefont {Nadler}, \citenamefont {Warner},\ and\
  \citenamefont {Zucker}}]{Coifman_pnas_2005}%
  \BibitemOpen
  \bibfield  {author} {\bibinfo {author} {\bibfnamefont {R.~R.}\ \bibnamefont
  {Coifman}}, \bibinfo {author} {\bibfnamefont {S.}~\bibnamefont {Lafon}},
  \bibinfo {author} {\bibfnamefont {A.~B.}\ \bibnamefont {Lee}}, \bibinfo
  {author} {\bibfnamefont {M.}~\bibnamefont {Maggioni}}, \bibinfo {author}
  {\bibfnamefont {B.}~\bibnamefont {Nadler}}, \bibinfo {author} {\bibfnamefont
  {F.}~\bibnamefont {Warner}},\ and\ \bibinfo {author} {\bibfnamefont {S.~W.}\
  \bibnamefont {Zucker}},\ }\bibfield  {title} {\bibinfo {title} {Geometric
  diffusions as a tool for harmonic analysis and structure definition of data:
  Diffusion maps},\ }\href {https://doi.org/doi:10.1073/pnas.0500334102}
  {\bibfield  {journal} {\bibinfo  {journal} {Proceedings of the National
  Academy of Sciences}\ }\textbf {\bibinfo {volume} {102}},\ \bibinfo {pages}
  {7426} (\bibinfo {year} {2005})}\BibitemShut {NoStop}%
\bibitem [{\citenamefont {Coifman}\ and\ \citenamefont
  {Lafon}(2006)}]{Coifman_acha_2006}%
  \BibitemOpen
  \bibfield  {author} {\bibinfo {author} {\bibfnamefont {R.~R.}\ \bibnamefont
  {Coifman}}\ and\ \bibinfo {author} {\bibfnamefont {S.}~\bibnamefont
  {Lafon}},\ }\bibfield  {title} {\bibinfo {title} {Diffusion maps},\ }\href
  {https://doi.org/https://doi.org/10.1016/j.acha.2006.04.006} {\bibfield
  {journal} {\bibinfo  {journal} {Applied and Computational Harmonic Analysis}\
  }\textbf {\bibinfo {volume} {21}},\ \bibinfo {pages} {5} (\bibinfo {year}
  {2006})}\BibitemShut {NoStop}%
\bibitem [{SM()}]{SM}%
  \BibitemOpen
  \href@noop {} {}\bibinfo {note} {See Supplemental Material about method,
  eigenstates of the probability matrix, NHSE, additional symmetries, Riemann
  surface, dynamical behavior and parameters setting.}\BibitemShut {Stop}%
\bibitem [{\citenamefont {Zhang}\ \emph {et~al.}(2018)\citenamefont {Zhang},
  \citenamefont {Shen},\ and\ \citenamefont {Zhai}}]{ZhaiHui_prl_2018}%
  \BibitemOpen
  \bibfield  {author} {\bibinfo {author} {\bibfnamefont {P.}~\bibnamefont
  {Zhang}}, \bibinfo {author} {\bibfnamefont {H.}~\bibnamefont {Shen}},\ and\
  \bibinfo {author} {\bibfnamefont {H.}~\bibnamefont {Zhai}},\ }\bibfield
  {title} {\bibinfo {title} {Machine learning topological invariants with
  neural networks},\ }\href {https://doi.org/10.1103/PhysRevLett.120.066401}
  {\bibfield  {journal} {\bibinfo  {journal} {Physical Review Letters}\
  }\textbf {\bibinfo {volume} {120}},\ \bibinfo {pages} {066401} (\bibinfo
  {year} {2018})}\BibitemShut {NoStop}%
\bibitem [{\citenamefont {Gong}\ \emph {et~al.}(2018)\citenamefont {Gong},
  \citenamefont {Ashida}, \citenamefont {Kawabata}, \citenamefont {Takasan},
  \citenamefont {Higashikawa},\ and\ \citenamefont
  {Ueda}}]{GongZongping_prx_2018}%
  \BibitemOpen
  \bibfield  {author} {\bibinfo {author} {\bibfnamefont {Z.}~\bibnamefont
  {Gong}}, \bibinfo {author} {\bibfnamefont {Y.}~\bibnamefont {Ashida}},
  \bibinfo {author} {\bibfnamefont {K.}~\bibnamefont {Kawabata}}, \bibinfo
  {author} {\bibfnamefont {K.}~\bibnamefont {Takasan}}, \bibinfo {author}
  {\bibfnamefont {S.}~\bibnamefont {Higashikawa}},\ and\ \bibinfo {author}
  {\bibfnamefont {M.}~\bibnamefont {Ueda}},\ }\bibfield  {title} {\bibinfo
  {title} {Topological phases of non-hermitian systems},\ }\href
  {https://doi.org/10.1103/PhysRevX.8.031079} {\bibfield  {journal} {\bibinfo
  {journal} {Physical Review X}\ }\textbf {\bibinfo {volume} {8}},\ \bibinfo
  {pages} {031079} (\bibinfo {year} {2018})}\BibitemShut {NoStop}%
\bibitem [{\citenamefont {Kawabata}\ \emph {et~al.}(2019)\citenamefont
  {Kawabata}, \citenamefont {Higashikawa}, \citenamefont {Gong}, \citenamefont
  {Ashida},\ and\ \citenamefont {Ueda}}]{kawabata2019topological}%
  \BibitemOpen
  \bibfield  {author} {\bibinfo {author} {\bibfnamefont {K.}~\bibnamefont
  {Kawabata}}, \bibinfo {author} {\bibfnamefont {S.}~\bibnamefont
  {Higashikawa}}, \bibinfo {author} {\bibfnamefont {Z.}~\bibnamefont {Gong}},
  \bibinfo {author} {\bibfnamefont {Y.}~\bibnamefont {Ashida}},\ and\ \bibinfo
  {author} {\bibfnamefont {M.}~\bibnamefont {Ueda}},\ }\bibfield  {title}
  {\bibinfo {title} {Topological unification of time-reversal and particle-hole
  symmetries in non-hermitian physics},\ }\href
  {https://doi.org/10.1038/s41467-018-08254-y} {\bibfield  {journal} {\bibinfo
  {journal} {Nature communications}\ }\textbf {\bibinfo {volume} {10}},\
  \bibinfo {pages} {297} (\bibinfo {year} {2019})}\BibitemShut {NoStop}%
\bibitem [{\citenamefont {Wang}\ \emph {et~al.}(2023)\citenamefont {Wang},
  \citenamefont {Liang}, \citenamefont {Ma}, \citenamefont {Fan}, \citenamefont
  {Ma}, \citenamefont {Jafri}, \citenamefont {Yang}, \citenamefont {Zhang},
  \citenamefont {Wang}, \citenamefont {Guo}, \citenamefont {Dong},
  \citenamefont {Liu}, \citenamefont {Wang}, \citenamefont {Hong},
  \citenamefont {Zhang}, \citenamefont {Gu}, \citenamefont {Yi}, \citenamefont
  {Zhang}, \citenamefont {Lin}, \citenamefont {Chen}, \citenamefont {Huang},\
  and\ \citenamefont {Nan}}]{WangJing_NC_2023}%
  \BibitemOpen
  \bibfield  {author} {\bibinfo {author} {\bibfnamefont {J.}~\bibnamefont
  {Wang}}, \bibinfo {author} {\bibfnamefont {D.}~\bibnamefont {Liang}},
  \bibinfo {author} {\bibfnamefont {J.}~\bibnamefont {Ma}}, \bibinfo {author}
  {\bibfnamefont {Y.}~\bibnamefont {Fan}}, \bibinfo {author} {\bibfnamefont
  {J.}~\bibnamefont {Ma}}, \bibinfo {author} {\bibfnamefont {H.~M.}\
  \bibnamefont {Jafri}}, \bibinfo {author} {\bibfnamefont {H.}~\bibnamefont
  {Yang}}, \bibinfo {author} {\bibfnamefont {Q.}~\bibnamefont {Zhang}},
  \bibinfo {author} {\bibfnamefont {Y.}~\bibnamefont {Wang}}, \bibinfo {author}
  {\bibfnamefont {C.}~\bibnamefont {Guo}}, \bibinfo {author} {\bibfnamefont
  {S.}~\bibnamefont {Dong}}, \bibinfo {author} {\bibfnamefont {D.}~\bibnamefont
  {Liu}}, \bibinfo {author} {\bibfnamefont {X.}~\bibnamefont {Wang}}, \bibinfo
  {author} {\bibfnamefont {J.}~\bibnamefont {Hong}}, \bibinfo {author}
  {\bibfnamefont {N.}~\bibnamefont {Zhang}}, \bibinfo {author} {\bibfnamefont
  {L.}~\bibnamefont {Gu}}, \bibinfo {author} {\bibfnamefont {D.}~\bibnamefont
  {Yi}}, \bibinfo {author} {\bibfnamefont {J.}~\bibnamefont {Zhang}}, \bibinfo
  {author} {\bibfnamefont {Y.}~\bibnamefont {Lin}}, \bibinfo {author}
  {\bibfnamefont {L.-Q.}\ \bibnamefont {Chen}}, \bibinfo {author}
  {\bibfnamefont {H.}~\bibnamefont {Huang}},\ and\ \bibinfo {author}
  {\bibfnamefont {C.-W.}\ \bibnamefont {Nan}},\ }\bibfield  {title} {\bibinfo
  {title} {Polar solomon rings in ferroelectric nanocrystals},\ }\href
  {https://doi.org/10.1038/s41467-023-39668-y} {\bibfield  {journal} {\bibinfo
  {journal} {Nature Communications}\ }\textbf {\bibinfo {volume} {14}},\
  \bibinfo {pages} {3941} (\bibinfo {year} {2023})}\BibitemShut {NoStop}%
\bibitem [{\citenamefont {Qian}(2007)}]{Qian_Hong_ARPC_2007}%
  \BibitemOpen
  \bibfield  {author} {\bibinfo {author} {\bibfnamefont {H.}~\bibnamefont
  {Qian}},\ }\bibfield  {title} {\bibinfo {title} {Phosphorylation energy
  hypothesis: Open chemical systems and their biological functions},\ }\href
  {https://doi.org/10.1146/annurev.physchem.58.032806.104550} {\bibfield
  {journal} {\bibinfo  {journal} {Annual Review of Physical Chemistry}\
  }\textbf {\bibinfo {volume} {58}},\ \bibinfo {pages} {113} (\bibinfo {year}
  {2007})}\BibitemShut {NoStop}%
\bibitem [{\citenamefont {Astumian}(2011)}]{Astumian_ARB_2011}%
  \BibitemOpen
  \bibfield  {author} {\bibinfo {author} {\bibfnamefont {R.~D.}\ \bibnamefont
  {Astumian}},\ }\bibfield  {title} {\bibinfo {title} {Stochastic
  conformational pumping: A mechanism for free-energy transduction by
  molecules},\ }\href {https://doi.org/10.1146/annurev-biophys-042910-155355}
  {\bibfield  {journal} {\bibinfo  {journal} {Annual Review of Biophysics}\
  }\textbf {\bibinfo {volume} {40}},\ \bibinfo {pages} {289} (\bibinfo {year}
  {2011})}\BibitemShut {NoStop}%
\bibitem [{\citenamefont {Utsumi}\ \emph {et~al.}(2010)\citenamefont {Utsumi},
  \citenamefont {Golubev}, \citenamefont {Marthaler}, \citenamefont {Saito},
  \citenamefont {Fujisawa},\ and\ \citenamefont
  {Sch\"on}}]{PhysRevB.81.125331}%
  \BibitemOpen
  \bibfield  {author} {\bibinfo {author} {\bibfnamefont {Y.}~\bibnamefont
  {Utsumi}}, \bibinfo {author} {\bibfnamefont {D.~S.}\ \bibnamefont {Golubev}},
  \bibinfo {author} {\bibfnamefont {M.}~\bibnamefont {Marthaler}}, \bibinfo
  {author} {\bibfnamefont {K.}~\bibnamefont {Saito}}, \bibinfo {author}
  {\bibfnamefont {T.}~\bibnamefont {Fujisawa}},\ and\ \bibinfo {author}
  {\bibfnamefont {G.}~\bibnamefont {Sch\"on}},\ }\bibfield  {title} {\bibinfo
  {title} {Bidirectional single-electron counting and the fluctuation
  theorem},\ }\href {https://doi.org/10.1103/PhysRevB.81.125331} {\bibfield
  {journal} {\bibinfo  {journal} {Physical Review B}\ }\textbf {\bibinfo
  {volume} {81}},\ \bibinfo {pages} {125331} (\bibinfo {year}
  {2010})}\BibitemShut {NoStop}%
\bibitem [{\citenamefont {Cao}\ \emph {et~al.}(2023)\citenamefont {Cao},
  \citenamefont {Li}, \citenamefont {Zhao}, \citenamefont {Guo}, \citenamefont
  {Qi}, \citenamefont {Chang}, \citenamefont {Zhou}, \citenamefont {Xu},\ and\
  \citenamefont {Duan}}]{PhysRevLett.130.163001}%
  \BibitemOpen
  \bibfield  {author} {\bibinfo {author} {\bibfnamefont {M.-M.}\ \bibnamefont
  {Cao}}, \bibinfo {author} {\bibfnamefont {K.}~\bibnamefont {Li}}, \bibinfo
  {author} {\bibfnamefont {W.-D.}\ \bibnamefont {Zhao}}, \bibinfo {author}
  {\bibfnamefont {W.-X.}\ \bibnamefont {Guo}}, \bibinfo {author} {\bibfnamefont
  {B.-X.}\ \bibnamefont {Qi}}, \bibinfo {author} {\bibfnamefont {X.-Y.}\
  \bibnamefont {Chang}}, \bibinfo {author} {\bibfnamefont {Z.-C.}\ \bibnamefont
  {Zhou}}, \bibinfo {author} {\bibfnamefont {Y.}~\bibnamefont {Xu}},\ and\
  \bibinfo {author} {\bibfnamefont {L.-M.}\ \bibnamefont {Duan}},\ }\bibfield
  {title} {\bibinfo {title} {Probing complex-energy topology via non-hermitian
  absorption spectroscopy in a trapped ion simulator},\ }\href
  {https://doi.org/10.1103/PhysRevLett.130.163001} {\bibfield  {journal}
  {\bibinfo  {journal} {Physical Review Letters}\ }\textbf {\bibinfo {volume}
  {130}},\ \bibinfo {pages} {163001} (\bibinfo {year} {2023})}\BibitemShut
  {NoStop}%
\end{thebibliography}%

\end{document}